\documentclass[12pt]{article}
\usepackage{epsfig,amssymb,amsmath,psfrag}
\usepackage[bulletsep]{collref}

\newcommand{\cN}{{\cal N}}

\newcommand{\cO}{{\cal O}}

\def \be  {\begin{equation}}
\def \ee  {\end{equation}}
\def \ba  {\begin{eqnarray}}
\def \ea  {\end{eqnarray}}

\newcommand \Li{{\rm Li}}

\newcommand \tbr [1] {\langle{#1}\rangle}
\newcommand \fbr [1] {({#1})}
\newcommand \mfbr [1] {\langle{#1}\rangle}

\newcommand{\mi}{{\rm\rule[2.4pt]{6pt}{0.65pt}}}
\newcommand{\pl}{\hspace{0.5pt}\text{{\small+}}\hspace{-0.5pt}}

\newcommand{\smallminus}{{\rm\rule[2.4pt]{6pt}{0.65pt}}}
\newcommand{\smallplus}{\hspace{0.5pt}\text{{\small+}}\hspace{-0.5pt}}
\newcommand \num [2] {({#1}\smallminus1\,{#1}\,{#1}\smallplus1)\cap({#2}\smallminus1\,{#2}\,{#2}\smallplus1)}

\begin{document}
\thispagestyle{empty}

\begin{flushright}
HU-EP-10/56\\
CERN-PH-TH/2010-237\\
LAPTH-042/10
\end{flushright}

\begingroup\centering
{\Large\bfseries\mathversion{bold} New differential equations for on-shell loop integrals \par}%
\vspace{7mm}

\begingroup\scshape\large
James M.~Drummond,
\endgroup
\vspace{3mm}

\begingroup
\textit{PH-TH Division, CERN, Geneva, Switzerland 
 }\\
and\\
\textit{LAPTH, Universit\'e de Savoie, CNRS\\
B.P. 110, F-74941 Annecy-le-Vieux Cedex, France}
\par
\texttt{drummond@lapp.in2p3.fr\phantom{\ldots}}
\endgroup

\vspace{0.5cm}

\begingroup\scshape\large
Johannes M.~Henn,
\endgroup
\vspace{3mm}

\begingroup
\textit{Institut f\"ur Physik, Humboldt-Universit\"at zu Berlin, \\
Newtonstra{\ss}e 15, D-12489 Berlin, Germany}\par
\texttt{henn@physik.hu-berlin.de\phantom{\ldots}}
\endgroup

\vspace{0.5cm}

\begingroup\scshape\large
Jaroslav Trnka,
\endgroup
\vspace{3mm}

\begingroup
\textit{School of Natural Sciences, Institute for Advanced Study, \\
and\\
 Department of Physics, Princeton University\\
Princeton, NJ 08540, USA }\par
\texttt{jtrnka@ias.edu\phantom{\ldots}}
\endgroup

\vspace{0.5cm}

\textbf{Abstract}\vspace{5mm}\par
\begin{minipage}{14.7cm}
We present a novel type of differential equations
for on-shell loop integrals.
The equations are second-order and
importantly, they reduce the loop level by one,
so that they can be solved
iteratively in the loop order.
We present several infinite series of integrals satisfying such
iterative differential equations.
The differential operators we use are best 
written using momentum twistor space.
The use of the latter was advocated in recent
papers discussing loop integrals in $\cN=4$ super Yang-Mills.
One of our motivations is to provide a tool for deriving analytical
results for scattering amplitudes in this theory.
We show that the 
integrals
needed for planar MHV amplitudes up to two loops 
can be thought of as deriving from a single master topology.
The master integral satisfies our differential equations, and so do most of the reduced
integrals.
A consequence of the differential equations is that the integrals we discuss are 
not arbitrarily complicated transcendental functions. For 
two
specific two-loop integrals 
we give the full analytic solution. The simplicity of the integrals appearing in the 
scattering amplitudes in planar $\mathcal{N}=4$ super Yang-Mills is strongly suggestive 
of a relation to the conjectured underlying integrability of the theory.
We expect these differential equations to be relevant
for all planar MHV and non-MHV amplitudes.
We also discuss possible extensions of our method
to more general classes of integrals.
\end{minipage}\par
\endgroup

\newpage

\tableofcontents

\setcounter{tocdepth}{2}

\newpage

\section{Introduction}

In this paper we present new differential equations for 
on-shell loop integrals. Our main motivation is to develop efficient
methods to determine the loop-level S-matrix in $\cN=4$ super
Yang-Mills (SYM), which is built from precisely the types
of integrals we discuss here. 
The planar $\mathcal{N}=4$ theory is believed to be governed by some underlying integrable structure and hence it certainly deserves to have a beautiful and simple S-matrix.
Our differential equations can be viewed as a step towards explaining this simplicity, as they
restrict the integrals contributing to the amplitudes. Our results are at the level of specific integrals and so can also be used for subsets of integrals appearing in less supersymmetric theories.
We also expect our method to be applicable to larger classes of integrals.
\\

The simplicity of the loop integrals implied by the differential equations we find is suggestive of a connection to the expected underlying integrability of planar $\cN=4$
SYM, and more concretely, of the underlying symmetries of
its tree-level amplitudes. 
It was found that the tree-level
S-matrix in this theory has a hidden symmetry, dual superconformal
symmetry \cite{Drummond:2008vq,Brandhuber:2008pf,Drummond:2008cr}.
Together with the ordinary superconformal symmetry
of the theory, it leads to a Yangian symmetry \cite{Drummond:2009fd}.
At loop level, {\it a priori} the symmetries are broken.\\

The breaking of the bosonic part of dual superconformal symmetry is well understood 
and is controlled by anomalous Ward identities \cite{Drummond:2007au,Drummond:2008vq}
that were initially derived for certain Wilson loops, and are conjectured to hold for the
scattering amplitudes. The Ward identities completely fix the functional form
of the four-point and five-point amplitudes, confirming the ABDK/BDS conjecture for these
cases \cite{Anastasiou:2003kj,Bern:2005iz}. Starting from six points, a modification is required, called 
the remainder function \cite{Drummond:2008aq,Bern:2008ap}.
An alternative way of understanding the dual conformal symmetry at loop level
is possible by using the massive regulator of \cite{Alday:2009zm}, which is inspired by the AdS/CFT correspondence \cite{Alday:2007hr}.
This setup is expected to make dual conformal symmetry exact, i.e. unbroken, at loop level.
Given this, it is natural to assume that the loop-level integral basis of $\cN=4$ SYM should
consist of integrals having this exact symmetry (see \cite{Henn:2010bk} and references therein).
Indeed, the absence of triangle sub-graphs at one loop was confirmed in \cite{Boels:2010mj} and 
further support for this conjecture comes from \cite{Bern:2010qa}. \\

A similar understanding of the full Yangian symmetry at loop
level is still lacking and subject to ongoing research \cite{Sever:2009aa,Bargheer:2009qu,Korchemsky:2009hm,Beisert:2010gn}.
Recently it was argued that the (unregulated)
loop {\it integrand} has the full Yangian symmetry, up to total derivatives \cite{ArkaniHamed:2010kv} . 
Indeed the integrand can be recursively constructed via a BCFW type recursion relation \cite{ArkaniHamed:2010kv} (see also \cite{Boels:2010nw}). 
Ignoring regularisation, this recursive construction can be written as a sequence of Yangian invariant operations on the basic Yangian 
invariant functions \cite{ArkaniHamed:2010kv} (see also \cite{ArkaniHamed:2009vw,Mason:2009qx,Drummond:2010uq,Korchemsky:2010ut} for a discussion of Yangian invariants).\\

Just as the dual conformal symmetry of the integrand \cite{Drummond:2006rz}
was a hint that there is an anomalous Ward identity \cite{Drummond:2007au},
the existence of the Yangian invariant integrand indicates that there
should be a way to directly understand the breaking of the full symmetry.
Although we do not yet make a direct connection to
the Yangian generators, we find it likely that the underlying Yangian
symmetry is 
related to
the differential equations we find.\\

Further very concrete motivation for our study also comes from recent explicit results for the 
hexagonal light-like Wilson loop, which is dual to the six-gluon MHV amplitude \cite{Drummond:2008aq,Bern:2008ap}
(for reviews see \cite{Alday:2008yw,Henn:2009bd} and references therein, and \cite{Mason:2010yk,CaronHuot:2010ek}
for recent developments).
In \cite{Goncharov:2010jf}, a remarkably simple form of the six-point remainder function  \cite{Drummond:2008aq,Bern:2008ap}
was given, based on previous work \cite{DelDuca:2010zg}.
Simple results are also conjectured to hold in special kinematics \cite{DelDuca:2010zp,Heslop:2010kq}. 
Recently, the analytic six-point Wilson loop result of \cite{DelDuca:2010zg}
could be reproduced in a kinematical
limit by an analytic calculation of the corresponding scattering amplitude,
extending previous calculations for mass-regulated amplitudes \cite{Henn:2010bk,Henn:2010ir}.
Related recent work on loop amplitudes
in $\cN=4$ SYM can be found in \cite{Kosower:2010yk,Alday:2010jz}.
See also \cite{Alday:2010zy,Eden:2010ce} for other related recent developments.\\

The use of differential equations to evaluate loop integrals \cite{Kotikov:1990kg,Kotikov:1991pm,Gehrmann:1999as}
is widespread in the literature,
for a review see chapter 7 of \cite{smirnov2006feynman}.
In this approach one considers a set of master integrals,
and an associated family of integrals is
be obtained from  by
shrinking lines.
One differentiates the master integrals
with respect to kinematical invariants or masses.
The result is in general a linear combination
of several integrals. Integral
reduction identities are then applied to re-express the latter
in terms of the set of master integrals.
In complicated cases this step can be non-trivial,
as in general it requires the knowledge of all reduction
identities.
In general one obtains a set of first-order differential equations
for the master integrals.
A disadvantage is that in
order to solve for a given integral, one has to deal with all
integrals of a given family. It can also happen
that when computing a finite integral,
intermediate steps contain
divergences that only cancel at the end,
and a regulator has to be used.
\\

The differential equations we obtain here are quite
different in nature. There are two important differences
to the method described above:
Firstly,
our differential equations can be applied directly to a given
integral, without having to know all integrals with
fewer propagators, or any integral reduction identities.
Secondly, they are second-order
equations, and, importantly, they reduce the loop level by one. In other
words, the homogeneous term of the equations
corresponds to lower-loop integrals, and the equations
therefore have an iterative structure.\\

While the differential equations method known in the
literature can be applied in principle to all loop integrals, the
differential equations we find are specific to a certain class
of integrals. The latter constitute a subset of integrals
needed for computing scattering amplitudes in a generic theory.
There is reason to believe that in the special case of $\cN=4$ SYM,
all planar loop integrals are constrained by the type of differential
equations we find here.
Very concretely we observe that the one- and two-loop MHV
amplitudes in $\cN=4$ SYM in the form given in \cite{ArkaniHamed:2010kv},
can all be expressed in terms of a single `master' integral, namely
\begin{equation}\label{eq-pictures-1}
\hspace{-4cm}
{
 \psfrag{oo}[cc][cc]{${}$}
\parbox[c]{30mm}{\includegraphics[height = 25mm]{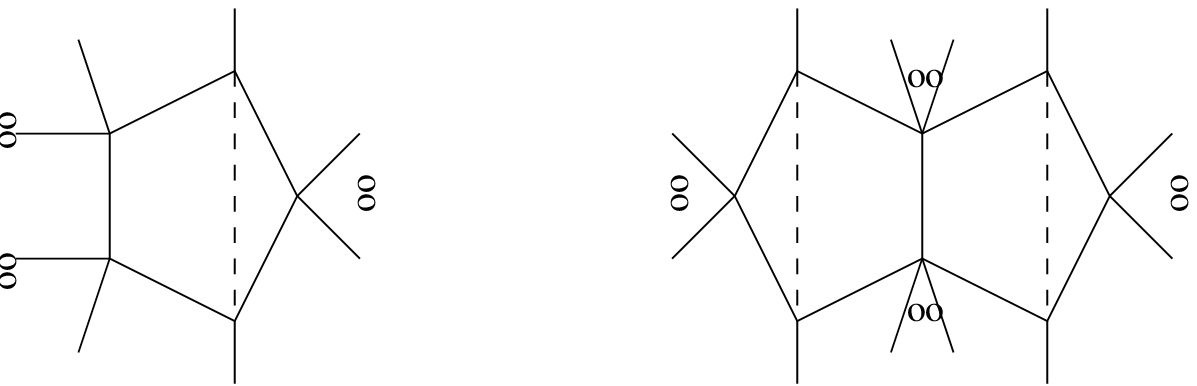}}
}
\end{equation}
at one- and two-loop order, respectively.
The reason is that reduced topologies, such as box, double box
or penta-box topologies, can be obtained by taking soft limits
of the master integrals.
We will give the precise definition of the integrals later.
In the six-point case, they were used recently to obtain an
analytic result for the remainder function at two loops (in a kinematical
limit) by two of the present authors in \cite{Drummond:2010mb}.
This is the first time that such an analytical result was obtained directly
for the scattering amplitude, and the results also hinted at a
certain simplicity of the integrals that had to be computed.
We will make this more precise in this paper.
As we will show, the integrals depicted in (\ref{eq-pictures-1})
satisfy second-order differential equations that reduce
their loop degree by one,
namely
\begin{equation}\label{eq-pictures-2}
\mathcal{D}^{(2)}\qquad
{
 \psfrag{oo}[cc][cc]{${}$}
\parbox[c]{30mm}{\includegraphics[height = 25mm]{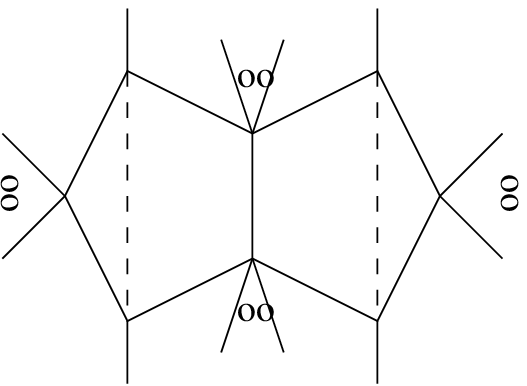}}
}
\qquad = \qquad
{
 \psfrag{oo}[cc][cc]{${}$}
\parbox[c]{30mm}{\includegraphics[height = 25mm]{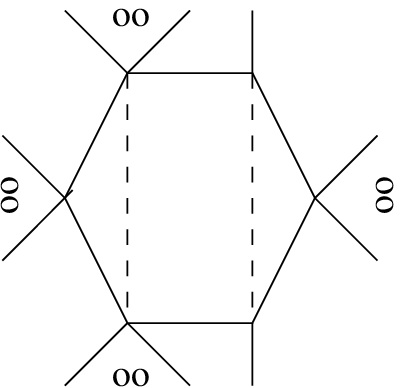}}
}
\end{equation}
The one-loop integrals appearing in (\ref{eq-pictures-1}) and on the r.h.s. of (\ref{eq-pictures-2})
satisfies a similar equation with simple rational functions as inhomogeneous terms.
These equations imply powerful constraints on the
functional dependence of those integrals.
In the following we will give several examples where
we solve such equations and obtain analytic
answers for two-loop integrals with multiple external legs.
\\

Iterative differential equations of the type (\ref{eq-pictures-2}) were previously found
in \cite{Drummond:2006rz} by two of the present authors and Smirnov and Sokatchev
for certain off-shell ladder (i.e. multiple box) integrals, and are closely
related to the equations we find here. We will review the differential equations
of \cite{Drummond:2006rz} in section \ref{sect-diffeq}.
The differential equations presented here can be thought of
as a generalisation of the ones used in \cite{Drummond:2006rz} to on-shell integrals.
\\

We find it very convenient to work with momentum twistor variables \cite{Hodges:2009hk}. These are well-suited to
describe planar loop integrals. They solve the momentum conservation and on-shell
constraints and therefore are unconstrained variables. This makes them the
natural variables to use if one is interested in differential equations.
One-loop box integrals were discussed in momentum
twistor space in \cite{Hodges:2010kq,Mason:2010pg}.
In the cases where infrared divergences are present we employ the
AdS-inspired mass regulator of \cite{Alday:2009zm}. The latter allows us to stay in
four dimensions and continue to use momentum twistors in those cases as well.
For recent references using momentum twistors and this regulator see  \cite{Hodges:2010kq,Mason:2010pg,Drummond:2010mb}.\\

The outline of the paper is as follows.
In section \ref{sect-moti}, we recall the definition of momentum
twistor variables and the expression for the one- and two-loop MHV
amplitudes in $\cN=4$ SYM. It is shown that at each loop level
there is only one `master' topology.
In section \ref{sect-diffeq} we show that certain classes of integrals
satisfy second-order differential equations that reduce the loop order
by one. We first review the differential equations for off-shell integrals
found in \cite{Drummond:2006rz}, and then generalise them to the on-shell case.
We present two mechanisms for finding such differential equations.
We present several infinite classes of integrals satisfying
iterative differential equations. In section \ref{sect-solve}
we give an example for how to solve the differential equations,
using certain assumptions about boundary conditions.
We give explicit analytical results
for several multi-leg integrals at two loops.
In section \ref{sect-outlook} we conclude and comment
on several possibilities of extending our method.
There are two Appendices. In Appendix A we discuss 
twistor differential operators that annihilate the integrand
of the integrals appearing in the MHV amplitudes, up to anomalies.
Appendix B contains the analytic formulas for the one- and two-loop 
penta-box integrals with ``magic'' numerator.

\section{Motivation: MHV amplitudes}
\label{sect-moti}

In this section we recall how momentum twistor variables
can be used to describe loop integrals. We also recall
the recently-proposed integral representation for planar
two-loop MHV amplitudes in $\cN=4$ super Yang-Mills \cite{ArkaniHamed:2010kv}.
We show that at each loop order, all integrals contained in
the MHV amplitudes can be thought of as deriving from
a single master topology. Reduced topologies are obtained
by taking soft limits. This is related to the consistency of the
loop integrand with soft limits  \cite{Drummond:2010mb}.\\

Let us briefly recall how the momentum twistor variables 
are related to the standard momentum
space variables. Given the $n$ incoming light-like momenta of a
planar colour-orderd ordered amplitude, \be p_i^{\alpha \dot \alpha} =
\lambda_i^\alpha \tilde{\lambda}_i^{\dot \alpha}\,, \ee we define
the dual coordinates in the usual manner
\cite{Broadhurst:1993ib,Drummond:2006rz}, \be x_i^{\alpha
\dot\alpha} - x_{i+1}^{\alpha \dot\alpha} = p_i^{\alpha
\dot\alpha}\,. \ee The dual $x_i$ define a light-like polygon in the
dual space. On the other hand, a point in dual coordinate space
corresponds to a (complex, projective) line in momentum twistor
space. Two dual points are light-like separated if the corresponding
lines in momentum twistor space intersect at some point in momentum
twistor space. Thus the light-like polygon in dual space corresponds
to a polygon in momentum twistor space with each line intersecting
its two neighbouring lines as each dual point is light-like
separated from its two neighbours. The $n$ momentum twistors
associated to this configuration of $n$ light-like lines are defined
via the incidence relations, \be \label{incidence} Z_i^A =
(\lambda_i^\alpha , \mu_i^{\dot\alpha}), \qquad \mu_i^{\dot\alpha} =
x_i^{\alpha \dot\alpha} \lambda_{i \alpha} = x_{i+1}^{\alpha
\dot\alpha} \lambda_{i \alpha}\,. \ee The momentum twistor
transforms linearly under the action of dual conformal symmetry, as
indicated by the fundamental $sl(4)$ index $A$. Moreover the $n$
momentum twistors describing the polygon are free variables, in
contrast to the dual points $x_i$ which obey the constraints of
light-like separation from their neighbours. The dual point $x_i$ is
associated to the line described by the pair $(Z_{i-1} Z_i)$ or $(i-1
\,\, i)$ for short. The incidence relations (\ref{incidence}) allow
one to express functions of the $x_i$ in terms of momentum twistors.
For example we have
\be x_{ij}^2 = \frac{\fbr{i\mi1\,i\,j\mi1\,j}}{\langle i\mi1\,i
\rangle \langle j\mi1\,j\rangle}\,, \ee
where the four-brackets and two-brackets are defined as follows,
\be \fbr{ijkl}  = \epsilon_{ABCD} Z_i^A Z_j^B Z_k^C Z_l^D, \qquad
\langle i j \rangle = \lambda_i^\alpha \lambda_{j \alpha}\,. \ee
The four-brackets are obviously dual conformal invariants while the two-brackets are invariant under just the Lorentz and (dual) translation transformations.\\

 \begin{figure}[t]
 \psfrag{dots}[cc][cc]{$\ldots$}
 \psfrag{z1}[cc][cc]{$Z_{3}$}
 \psfrag{z2}[cc][cc]{$Z_{4}$}
 \psfrag{zn}[cc][cc]{$Z_{2}$}
 \psfrag{znm}[cc][cc]{$Z_{1}$}
 \psfrag{zi}[cc][cc]{$Z_{i}$}
 \psfrag{zim}[cc][cc]{$Z_{i-1}$}
 \psfrag{zab}[cc][cc]{$Z_{A}Z_{B}$}
 \psfrag{x0}[cc][cc]{$x_{0}$}
 \psfrag{xn}[cc][cc]{$x_{2}$}
 \psfrag{x1}[cc][cc]{$x_{3}$}
 \psfrag{x2}[cc][cc]{$x_{4}$}
 \psfrag{xi}[cc][cc]{$x_{i}$}
  \psfrag{k}[cc][cc]{$k$}
  \psfrag{p1}[cc][cc]{$p_{3}$}
  \psfrag{p2}[cc][cc]{$p_{4}$}
  \psfrag{pim}[cc][cc]{$p_{i-1}$}
 \centerline{
 {\epsfxsize13cm  \epsfbox{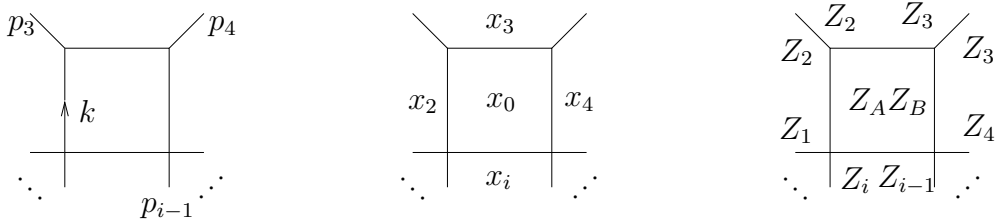}}
}
\caption{\small
`Two-mass-hard' integral in momentum space, dual space, and momentum twistor space variables, see equations (\ref{twistorbox}) and (\ref{eq-2mh1loop}).}
\label{fig-2mh}
\end{figure}

We can consider integrals over the space of lines in momentum twistor space $(AB)$. As we have discussed, this is equivalent to an integration over points in dual space.
For example, the two-mass hard integral of Fig. \ref{fig-2mh} is given by
\begin{equation} \label{twistorbox}
\int \frac{d^{4}k}{i \pi^2}  \, \frac{ (p_{3}+ \ldots + p_{i-1})^2 (p_{2}+p_{3})^2}{k^2 (k+p_{2})^2 (k+p_{2}+p_{3})^2 (k+p_{2}+ \ldots + p_{i-1})^2 }= \int \frac{ d^{4}x_{0} }{i \pi^2} \, \frac{ x_{3i}^2 x_{24}^2} { x_{02}^2 x_{03}^2 x_{04}^2 x_{0i}^2} \,,
\end{equation}
where the first expression is written in momentum space and the second one in dual coordinates.
In twistor notation, this becomes
\begin{equation}\label{eq-2mh1loop}
\int \frac{d^4Z_{AB}}{i \pi^2} \,  \frac{\fbr{23\,i\mi1 \, i} \fbr{1
2 3 4 }}{  \fbr{AB12} \fbr{AB 2 3 }\fbr{AB 3 4}\fbr{AB\, i\mi1\, i}
} \,.
\end{equation}
The integral (\ref{twistorbox}) as written is infrared divergent, so
its proper definition should involve a regulator. We will use the
the AdS regularisation introduced in \cite{Alday:2009zm}. The latter
allows us to stay in four dimensions and regulates the infrared
divergences by masses that in turn are generated by a Higgs
mechanism. For actual calculations we will use the regularisation
where all masses are equal. In particular this means the outermost
propagators in the planar loop integrals that we are studying are
modified as follows,

\be \frac{1}{x_{ij}^2} \longrightarrow \frac{1}{x_{ij}^2 + m^2}\,.
\ee
In twistor space this has the effect that each of the propagator
factors becomes

\be \frac{1}{ \fbr{AB\,i\mi1\,i}} \longrightarrow \frac{1}{\mfbr{ AB
\,i\mi1 \, i}} \equiv   \frac{1}{\fbr{AB\, i\mi1 \,i} + m^2 \tbr{ AB
} \tbr{ i\mi1 \, i}}\,. \ee
When doing this, there is a choice of adding $\cO(m^2)$ terms to the
integrand. This is certainly relevant if one wishes to obtain the
exact $m$-dependence of amplitudes on the Coulomb branch of $\cN=4$
SYM. In the cases considered here we are mostly interested in the
case where $m$ is small. Unless an integral diverges linearly as $m
\to 0$ one can
then drop such numerator terms. This is the case for all integrals considered here.\\

\begin{figure}[t]
\psfrag{dots}[cc][cc]{$\ldots$}
 \psfrag{z1}[cc][cc]{${}$}
 \psfrag{z2}[cc][cc]{${}$}
 \psfrag{zn}[cc][cc]{${}$}
 \psfrag{znm}[cc][cc]{${}$}
 \psfrag{zi}[cc][cc]{${}$}
 \psfrag{zim}[cc][cc]{${}$}
   \psfrag{i}[cc][cc]{$x_{i}$}
  \psfrag{ip}[cc][cc]{$x_{i+1}$}
  \psfrag{ipp}[cc][cc]{{$x_{i+2}$}}
  \psfrag{j}[cc][cc]{$x_{j}$}
  \psfrag{jpp}[cc][cc]{{$x_{j+2}$}}
  \psfrag{jp}[cc][cc]{$x_{j+1}$}
  \psfrag{k}[cc][cc]{$x_{k}$}
  \psfrag{kp}[cc][cc]{$x_{k+1}$}
  \psfrag{kpp}[cc][cc]{$x_{k+2}$}
  \psfrag{l}[cc][cc]{$x_{\ell}$}
  \psfrag{lp}[cc][cc]{$x_{\ell+1}$}
   \psfrag{a}[cc][cc]{$I^{\rm pent}_{n;i,j,k}$}
   \psfrag{b}[cc][cc]{$I^{\rm box}_{n;i,j}$}
 \centerline{
 {\epsfxsize9cm  \epsfbox{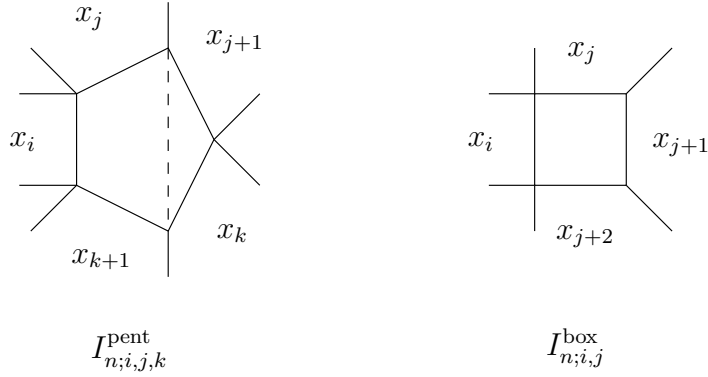}}
}\caption{\small Integrals contributing to the one-loop MHV
amplitude. The box integrand is obtained from the pentagon integrand
by taking the soft limit $p_{n} \to 0$. 
The
dashed line represents here the numerator which is denoted by wiggly
line in \cite{ArkaniHamed:2010kv}.} \label{fig-mhv1loop}
\end{figure}

Let us now discuss how such integrals appear in scattering
amplitudes in $\cN=4$ SYM. At one loop, the MHV amplitudes are
usually represented (or more precisely, their parity-even part) as a
sum over so-called two-mass easy box integrals \cite{Bern:1994zx}. In
\cite{ArkaniHamed:2010kv}, an alternative form was given that uses
pentagon integrals with certain twistor numerators. The formula
given in \cite{ArkaniHamed:2010kv} is a sum over the integrals shown
in Fig. \ref{fig-mhv1loop}.
Their integrand is given by

\begin{eqnarray}\label{eq1loopmhv}
I^{\rm pent}_{n;i,j,k} &=& \frac{ \fbr{i\mi1 \,i\,j\,k} \fbr{AB
\num{j}{k}}}{ \fbr{AB\, i\mi1\, i} \fbr{AB\, j\mi1\, j} \fbr{AB j \,
j\pl1 }\fbr{AB\, k\mi1\, k }\fbr{AB k\, k\pl1 } } \,.
\end{eqnarray}
We use the notation $I$ for the integrand and $F$ for the function obtained after integration\footnote{As was mentioned before, some
of the integrals are in fact infrared divergent and require a
regulator. In this section we mostly discuss properties of the
(unregulated) loop integrand. It was argued in
\cite{ArkaniHamed:2010kv} that a correct expression for the loop
integrand on the Coulomb branch of $\cN=4$ SYM \cite{Alday:2009zm}, up to $\cO(m^2)$
terms, can be obtained by adding masses to propagators on the
perimeter of the diagams.}, i.e.
\begin{eqnarray}
F^{\rm pent}_{n;i,j,k} &=&  \int \frac{d^{4}Z_{AB}}{i \pi^2} \, I^{\rm pent}_{n;i,j,k} \,,
\end{eqnarray}
where the number of points $n$ is implicit and appears only in the condition $n+1 \equiv 1$.\\

A comment is in order here regarding the numerator factors: In
\cite{ArkaniHamed:2010kv} both one-loop and two-loop amplitudes are
written using numerators with wiggly lines representing
\be
\fbr{AB\num{i}{j}} := \fbr{A \, i\smallminus1\, i \, i \smallplus 1} \fbr{B\, j \, \smallminus 1 \,  j \, j\smallplus1} - \fbr{B\, i\smallminus1\, i \, i \smallplus 1} \fbr{A \, j\smallminus1\, j \, j\smallplus1}\,.
\ee
The $\overline{\mbox{MHV}}$ amplitude is
obtained by changing wiggly lines to dashed lines that stand for
$\fbr{AB\,i\,j}$, etc. In our discussion we do not distinguish
between an integral and the parity conjugate integral because their
difference is a parity odd integral which
integrates to $\cO(m^2)$.
Therefore, in any diagram we can change all (but not just some) wiggly lines to dashed lines
and vice-versa. In the following we will not distinguish between integrals with all
dashed lines and those with all wiggly lines.\\

We now argue that all integrals appearing in the one-loop MHV
amplitudes can be thought of as deriving from the pentagon integral.
It is clear that pentagon integrals with lower number of legs can be
obtained from a generic pentagon integral by taking soft limits.
Therefore we only have to show how to obtain the box integrals of
equation (\ref{eq-2mh1loop}). As we will see presently, they are
also obtained by taking soft limits. Indeed, consider the pentagon
integral of  Fig. \ref{fig-mhv1loop} with $k=j+2$, so that only one
external leg enters its rightmost corner. The soft limit of $p_{j+1}
\to 0$ corresponds to letting
\begin{equation}
Z_{j+1} \to \alpha Z_{j} + \beta Z_{j+2}\,.
\end{equation}
Applying this limit to the integrand of the pentagon integral given
in (\ref{eq1loopmhv}), we find
\begin{equation}
\lim_{p_{j+1} \to 0} I^{\rm pent}_{n;i,j,j+2} \to  I^{\rm
box}_{n-1;i,j} \,,
\end{equation}
one reproduces exactly the integrand of the ``two-mass hard'' box.
In this sense we can say that the one-loop
MHV amplitude is built from a single master integral.
We will show in section \ref{sect-diffeq} that the latter satisfies
a differential equation.\\

 \begin{figure}[t]
\psfrag{dots}[cc][cc]{$\ldots$}
 \psfrag{z1}[cc][cc]{$Z_{1}$}
 \psfrag{z2}[cc][cc]{$Z_{2}$}
 \psfrag{zn}[cc][cc]{$Z_{n}$}
 \psfrag{znm}[cc][cc]{$Z_{n-1}$}
 \psfrag{zi}[cc][cc]{$Z_{i}$}
 \psfrag{zim}[cc][cc]{$Z_{i-1}$}
  \psfrag{p1}[cc][cc]{$I^{\rm double-pent}_{n;i,j,k,\ell}$}
  \psfrag{p2}[cc][cc]{$I^{\rm penta-box}_{n;i,j,k}$}
  \psfrag{p3}[cc][cc]{$I^{\rm double-box}_{n;i,k}$}
    \psfrag{i}[cc][cc]{\scriptsize $x_i$}
  \psfrag{ip}[cc][cc]{\scriptsize $x_{i+1}$}
  \psfrag{ipp}[cc][cc]{\scriptsize{$x_{i+2}$}}
  \psfrag{j}[cc][cc]{\scriptsize $x_{j}$}
  \psfrag{jp}[cc][cc]{\scriptsize $x_{j+1}$}
  \psfrag{k}[cc][cc]{\scriptsize $x_{k}$}
  \psfrag{kp}[cc][cc]{\scriptsize $x_{k+1}$}
  \psfrag{kpp}[cc][cc]{\scriptsize $x_{k+2}$}
  \psfrag{l}[cc][cc]{\scriptsize $x_{\ell}$}
  \psfrag{lp}[cc][cc]{\scriptsize $x_{\ell+1}$}
 \centerline{
 {\epsfxsize15cm  \epsfbox{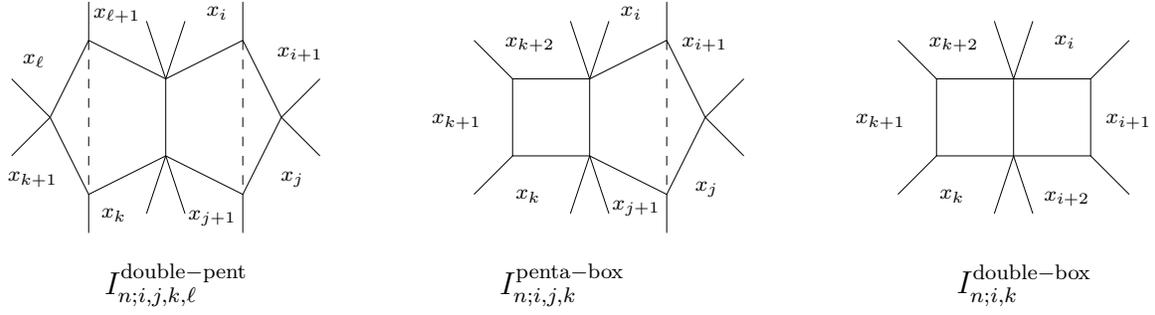}}
} \caption{\small Integrals contributing to the two-loop MHV
amplitude. The integrands of the double box and penta-box integrals
are obtained from the double pentagon integrals by taking the soft
limits. The dashed line have again the same meaning as in 1-loop
case.} \label{fig-mhv2loop}
\end{figure}

The situation is almost identical at the two-loop level.
The expression given in \cite{ArkaniHamed:2010kv}  involves
only the integrals shown in Fig  \ref{fig-mhv2loop}.
Recall that we denote the integrand of loop integrals by $I$ and the functions obtained after
integration by $F$. For example, the first integral shown in Fig. \ref{fig-mhv2loop}
has the following definition,
\begin{eqnarray}
I^{\rm double-pent}_{n;i,j,k,\ell} &=&  \frac{N}{\fbr{ABCD}}  \frac{\fbr{CD\num{i}{j}}}
{\fbr{CD \, i\smallminus1\,i} \fbr{CD \, i\,i\smallplus1} \fbr{CD\, j\smallminus1\,j} \fbr{CD\, j\,j\smallplus1} } \times \nonumber \\
&& \times   \frac{\fbr{AB \num{k}{\ell}} }{\fbr{AB \, k\smallminus1\,k}
\fbr{AB \,k\,k\smallplus1} \fbr{AB\, \ell\smallminus1\, \ell} \fbr{AB\, \ell \,
\ell\pl1} } \,,
\end{eqnarray}
with the normalisation $N=\fbr{i\,j\,k\,\ell}$ and
\begin{equation}\label{eq-F-double-pent}
F^{\rm double-pent}_{n;i,j,k,\ell} = \int \frac{ d^{4}Z_{AB}}{i \pi^2}\frac{ d^{4}Z_{CD}}{i \pi^2} \, I^{\rm double-pent}_{n;i,j,k,\ell}\,,
\end{equation}
where the number of points $n$ is implicit and appears only in the condition $n+1 \equiv 1$.
The double pentagon integrals defined in (\ref{eq-F-double-pent}) are in fact infrared finite \cite{ArkaniHamed:2010kv}.
Some of the pentagon and all of the box integrals we consider are infrared divergent and
their definition is understood with the mass regulator in place, which leads to the modifications discussed above.
{}From the one-loop case
it is clear that the penta-box and double box integrals shown
in Fig. \ref{fig-mhv2loop} can be obtained from the double pentagon integral
by taking subsequent soft limits.\\

So, in summary, at each loop level the integrals appearing
in the MHV amplitudes can be thought of as deriving from
a single master topology. We will show in section \ref{sect-diffeq}
that the latter integral satisfies a second-order differential
equation that reduces its loop order by one.\\

The relationship between the master integrals and the reduced
integrals works at the level of the integrand. When the integration
is taken into account, in some cases infrared divergences can
lead to a non-commutativity of the soft limit and the regulator limit.
We adopt the point of view that in those cases the relation to the
master integral still implies a certain simplicity of the reduced
integral. We show this explicitly for IR-divergent penta-box integrals
in section \ref{sect-diffeq}, which satisfy the same type of differential equations.\\

Let us briefly discuss how many functionally different integrals
of this class exist. The most general double pentagon integral
of the type shown in Fig. \ref{fig-mhv2loop} is the one where four of its
external legs are doubled (when more legs are added, the function
does not change). This is possible for the first time at $n=12$,
e.g. $F^{\rm double-pent}_{12;1,4,7,10}$.
In general, we can consider all integrals $F^{\rm double-pent}_{n;i,j,k,\ell}$
with zero, one, or two momenta flowing into these corners. Because of the symmetries
of the pentagon, there are $18$ different possibilities.
However, because of the finiteness of the double pentagon integrals,
all of them can be obtained from the most general case $F^{\rm double-pent}_{12;1,4,7,10}$ by
taking soft limits.
Finally, there are $12$ penta-box integrals $F^{\rm penta-box}_{n;i,j,k}$ and $6$ double
box integrals of the type $F^{\rm double-box}_{n;i,k}$.
The latter are all infrared divergent and therefore depend on the regulator $m^2$.

\section{Differential equations for loop integrals}
\label{sect-diffeq}

\subsection{Finite integrals}
\label{sect-diff-finite}

\subsubsection{Ladders}

The ladders (or scalar boxes) provide a first example of a class of integrals which satisfy the type of differential equations we are interested in. To understand the differential equations for the ladder integrals it is convenient to use the dual coordinate notation. For the integrals appearing in the two-loop amplitude discussed in section \ref{sect-moti} we will pass to the momentum twistor notation.
We will begin with the one-loop box as an example, using dual variables to express the momenta,
 \begin{figure}[t]
\psfrag{dots}[cc][cc]{$\ldots$}
 \psfrag{xi}[cc][cc]{$x_{i}$}
 \psfrag{xj}[cc][cc]{$x_{j}$}
 \psfrag{xk}[cc][cc]{$x_{k}$}
 \psfrag{xl}[cc][cc]{$x_{l}$}
   \centerline{
 { \epsfysize3cm \epsfbox{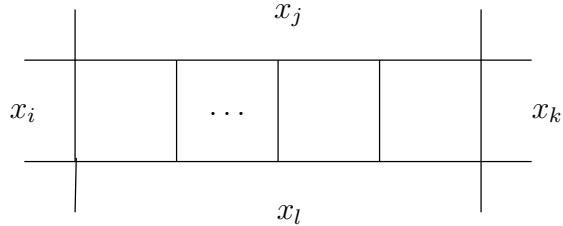} }
} \caption{\small 
Ladder integrals defined in equations (\ref{1box}) and (\ref{def-boxes}).} \label{fig-ladders}
\end{figure}
\be
\tilde{F}^{(1)}(x_i,x_j,x_k,x_l) = \int \frac{d^4 x_r}{i \pi^2} \frac{1}{x_{ir}^2 x_{jr}^2 x_{kr}^2 x_{lr}^2} \, .
\label{1box}
\ee
Note that the points $x_i,x_j,x_k,x_l$ are generic points without any light-like separations.
The $L$-loop version of this diagram is shown in Fig. \ref{fig-ladders}.
Since the integral (\ref{1box}) is covariant under conformal transformations of the $x$ coordinates, it can be expressed in terms of a function of the two available cross-ratios,
\be
u = \frac{x_{ij}^2 x_{kl}^2}{x_{ik}^2 x_{jl}^2}, \qquad v = \frac{x_{il}^2 x_{jk}^2}{x_{ik}^2 x_{jl}^2}\, .
\ee
Thus we have
\be
\tilde{F}^{(1)}(x_i,x_j,x_k,x_l) =\frac{\Phi^{(1)}(u,v)}{x_{ik}^2 x_{jl}^2} \,,
\label{I1explicit}
\ee
where the function $\Phi^{(1)}(u,v)$ is known \cite{Usyukina:1993ch,Broadhurst:1993ib,Isaev:2003tk}. In fact the function is best expressed in terms of the variables $z$ and $\bar{z}$ defined by\footnote{Note that $z$ and $\bar{z}$ are real and independent in Minkowski signature whereas they are complex conjugate to each other in Euclidean signature.}
\be
u = \frac{z \bar z}{(1-z)(1-\bar z)}
\,, \qquad
v = \frac{1}{(1-z)(1-\bar z)} \,.
\ee
Explicitly it is given by
\be
\Phi^{(1)}(u,v) = \frac{f^{(1)}(z,\bar{z})}{z-\bar{z}},\qquad f^{(1)}(z,\bar{z}) =  \log(z \bar{z})\bigl( \Li_1(z)-\Li_1(\bar{z}) \bigr)- 2\bigl(\Li_2(z) - \Li_2(\bar{z})\bigr)\,.
\ee
The integral $\tilde{F}^{(1)}$ satisfies a simple second order differential equation.
The reason for this is that acting with the Laplace operator on one of the external points, $x_i$ say,
produces a delta function under the integral \cite{Drummond:2006rz},
\be
\Box_i \frac{1}{x_{ir}^2} = -4 i \pi^2 \delta^{(4)}(x_i-x_r) \,.
\label{Laplacedelta}
\ee
This has the effect of localising the integral completely, giving a simple second-order equation,
\be
\Box_i   \tilde{F}^{(1)}(x_i,x_j,x_k,x_l)  = \frac{-4}{x_{ij}^2 x_{ik}^2 x_{il}^2} \, .
\label{boxI1int}
\ee
On the other hand \cite{Drummond:2006rz}, acting on the form of $\tilde{F}^{(1)}$ given in (\ref{I1explicit}) one obtains
\be
\Box_i \tilde{F}^{(1)}(x_i,x_j,x_k,x_l) = \frac{x_{jk}^2 x_{kl}^2}{x_{ik}^6 x_{jl}^4} \Delta_{u,v} \Phi^{(1)}(u,v)\,
\label{boxI1func}
\ee
where $\Delta_{u,v}$ is a second-order differential operator,
\be
\Delta_{u,v} = u\partial_u^2 + v \partial_v^2 + (u + v - 1)\partial_u \partial_v + 2 \partial_u + 2 \partial_v\,.
\ee
The equality of the two expressions (\ref{boxI1int}) and (\ref{boxI1func}) is a second order differential equation for the function $\Phi^{(1)}$ or equivalently $f^{(1)}$.
Expressing the equation in terms of $z$ and $\bar{z}$ it reads,
\be
z \partial_z \bar{z} \partial_{\bar{z}}  f^{(1)}(z, \bar{z})  = \frac{z}{z-1}-\frac{\bar{z}}{\bar{z}-1}\,.
\ee
The main points we wish to stress are that the action of the operator removes the loop integration, leaving a rational function behind. The existence of a simple equation also means that the underlying function is a relatively simple pure transcendental function of degree two.

The one-loop integral we have been discussing is just the first in an infinite sequence of ladder integrals. These integrals (along with a large class of equivalent integrals \cite{Drummond:2006rz}) exhibit an iterative structure,
\begin{align}\label{def-boxes}
 \tilde{F}^{(L)}(x_i,x_j,x_k,x_l) &= \int \frac{d^4 x_r}{i \pi^2} \frac{ x_{jl}^2}{x_{ir}^2 x_{jr}^2 x_{lr}^2} \tilde{F}^{(L-1)}(x_r,x_j,x_k,x_l)\,.
\end{align}
As before the integrals are invariant under conformal transformations of the $x_i$ and so are given in terms of functions of the the two cross-ratios,
\be
 \tilde{F}^{(L)}(x_i,x_j,x_k,x_l) = \frac{\Phi^{(L)}(u,v)}{x_{ik}^2 x_{jl}^2} \,.
\ee
As before the function $\Phi^{(n)}$ is best expressed in terms of the variables $z$ and $\bar{z}$,
\be
\Phi^{(L)}(u,v) = \frac{f^{(L)}(z,\bar{z})}{z-\bar{z}}\,,
\ee
where
\be
f^{(L)}(z,\bar{z}) = \sum_{r=0}^L \frac{(-1)^{r+L} (2L-r)!}{r! (L-r)! L!} \log^r (z \bar{z}) \bigl(\Li_{2L-r}(z) - \Li_{2L-r} (\bar{z})\bigr)\,.
\ee
The fact that the one-loop box satisfies a differential equation guarantees that all the ladder integrals do. One can see that Laplace operator $\Box_i$ is effectively acting only on a one-loop box subintegral which we have already seen reduces to a rational function. Thus the operation reduces the loop order of the ladder integral,
\be
\Box_i \tilde{F}^{(L)}(x_i,x_j,x_k,x_l) = -4  \frac{x_{jl}^2}{x_{ij}^2 x_{il}^2} \tilde{F}^{(L-1)}(x_i,x_j,x_k,x_l) \,.
\ee
The functions appearing in the explicit expressions exhibit corresponding differential equations
\be
z \partial_z \bar{z} \partial_{\bar{z}} f^{(L)}(z,\bar{z}) = f^{(L-1)}(z,\bar{z})\, .
\ee

\subsubsection{Pentaladders}

 \begin{figure}[t]
\psfrag{dots}[cc][cc]{$\ldots$}
 \psfrag{x1}[cc][cc]{$x_1$}
 \psfrag{x2}[cc][cc]{$x_2$}
 \psfrag{x3}[cc][cc]{$x_3$}
 \psfrag{x4}[cc][cc]{$x_4$}
 \psfrag{x5}[cc][cc]{$x_5$}
 \psfrag{x6}[cc][cc]{$x_6$}
  \psfrag{x7}[cc][cc]{$x_7$} 
   \psfrag{z1}[cc][cc]{$x_1$}
 \psfrag{z2}[cc][cc]{$x_2$}
 \psfrag{z3}[cc][cc]{$x_3$}
 \psfrag{z4}[cc][cc]{$x_4$}
 \psfrag{z5}[cc][cc]{$x_5$}
 \psfrag{z6}[cc][cc]{$x_6$}
  \psfrag{z7}[cc][cc]{$x_7$} 
    \psfrag{a}[cc][cc]{$(a)$} 
      \psfrag{b}[cc][cc]{$(b)$} 
  \centerline{
 { \epsfysize3cm \epsfbox{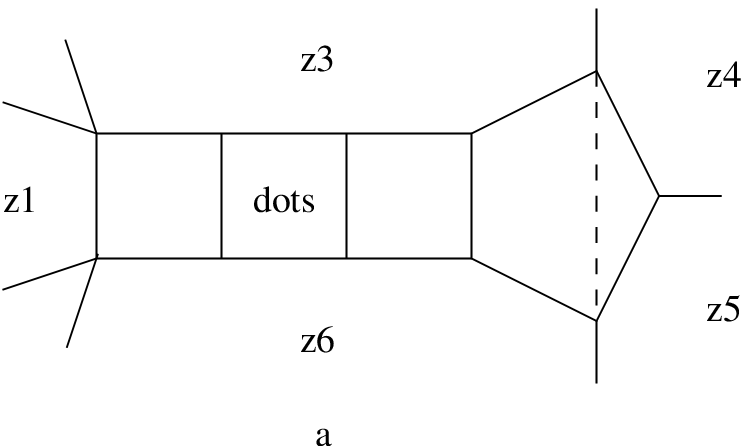}  \hspace{2cm} \epsfysize3cm \epsfbox{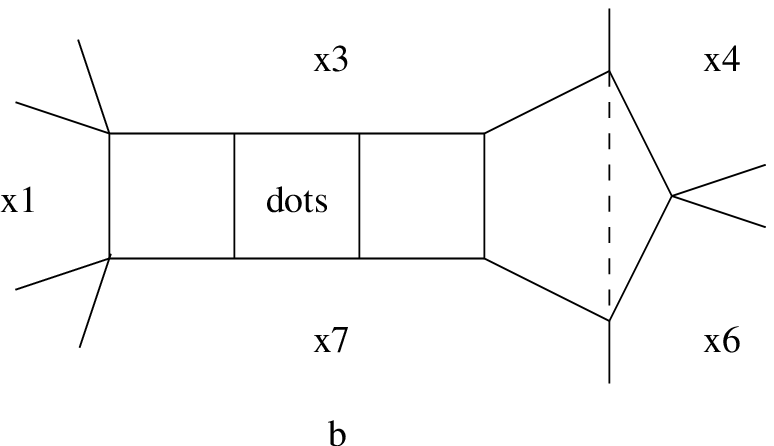} }
} \caption{\small 
Figure (a) represents the seven-point integrals defined in (\ref{1pent}) and (\ref{loopspent}),
and (b) shows the eight-point integrals defined in (\ref{1pentmassive}) and (\ref{looppentmassive}). } \label{fig-pentaladders}
\end{figure}

The integrals involving twistor numerators can also satisfy differential equations. We will begin with integrals which are similar to the ladders of the previous section. We would like to consider the finite pentaladder integrals, beginning with one of the one-loop pentagon integrals discussed in section \ref{sect-moti}. We will write in dual coordinate notation to begin with and later move to the momentum twistor notation. The integral we would like to consider is the seven-point one-loop pentagon integral, which is shown for arbitrary number of loops in Fig. \ref{fig-pentaladders}(a),
\be
\tilde{F}_{\rm pl}^{(1)}(x_1,x_3,x_{4},x_{5},x_{6}) = \int \frac{d^4 x_r}{i \pi^2} \frac{ x_{ar}^2}{x_{1r}^2 x_{3r}^2 x_{4r}^2 x_{5r}^2 x_{6r}^2} \,.
\label{1pent}
\ee
Note that we have not normalised the integral so (\ref{1pent}) is dimensionful.
The point $x_a$ is the magic complex point which is null-separated from every point on the null lines $(x_3, x_4),(x_{4}, x_{5})$ and $(x_{5}, x_{6})$. It is one of the solutions to
\be
x_{a3}^2=x_{a4}^2=x_{a5}^2=x_{a6}^2=0 \, .
\ee
The other solution is its parity conjugate.

Just like the ladder integrals, the pentagon we are considering is conformally covariant so we can write it in the following way
\be
\tilde{F}^{(1)}_{\rm pl}(x_1,x_3,x_{4},x_{5},x_{6}) = \frac{x_{1a}^2}{x_{14}^2 x_{15}^2 x_{36}^2}\frac{ \Psi^{(1)}(u,v)}{(1-u-v)} \,,
\label{I1}
\ee
where $u$ and $v$ are the two non-vanishing conformal cross-ratios,
\be
u = \frac{x_{13}^2 x_{46}^2}{x_{14}^2 x_{36}^2}, \qquad v = \frac{x_{16}^2 x_{35}^2}{x_{15}^2 x_{36}^2} \,.
\ee
We have chosen to make a factor of $1/(1-u-v)$ explicit on the RHS of (\ref{I1}) for later convenience.

In terms of the quantities introduced in section \ref{sect-moti} we have
\be
x_{14}^2 x_{15}^2 x_{36}^2 \tilde{F}^{(1)}_{\rm pl}(x_1,x_3,x_{4},x_{5},x_{6}) = \Psi^{(1)}(u,v) = F_{7;1,3,5}^{\rm pent}
\ee

The integral satisfies a differential equation very similar to the equation for the ladder integrals,
\be
\Box_1 \tilde{F}^{(1)}_{\rm pl} = -4 \frac{x_{1a}^2}{x_{13}^2 x_{14}^2 x_{15}^2 x_{16}^2} \,.
\ee
This equation translates into a second order p.d.e. for the function $\Psi^{(1)}$,
\be
uv \partial_u \partial_v \Psi^{(1)}(u,v) = 1\,.
\label{psi1pde}
\ee
Just as for the ladders we can define multi-loop pentaladder integrals via an iterative structure. For example we can consider the integrals, see Fig. \ref{fig-pentaladders}(a),
\be
\tilde{F}^{(L)}_{\rm pl}(x_1,x_3,x_{4},x_{5},x_{6}) = \int \frac{d^4x_r }{i \pi^2} \frac{ x_{36}^2}{x_{1r}^2 x_{3r}^2 x_{6r}^2} \tilde{F}^{(L-1)}_{\rm pl}(x_r,x_3,x_{4},x_{5},x_{6}) \,.
\label{loopspent}
\ee
Equivalently we can write them using the ladder integrals,
\be
\tilde{F}^{(L)}_{\rm pl}(x_1,x_3,x_{4},x_{5},x_{6}) = \int \frac{d^4x_r}{i \pi^2} \frac{ x_{ar}^2 x_{36}^2}{x_{3r}^2 x_{4r}^2 x_{5r}^2 x_{6r}^2} \tilde{F}^{(L-1)}(x_1,x_3,x_r,x_6)\,.
\ee

As before, we will use conformal symmetry in the $x$ variables to write the integral in the form
\be
\tilde{F}^{(L)}_{\rm pl}(x_1,x_3,x_{4},x_{5},x_{6}) = \frac{x_{1a}^2}{x_{14}^2 x_{15}^2 x_{36}^2} \frac{\Psi^{(L)}(u,v)}{(1-u-v)} \,.
\ee
Now applying the Laplace operator gives the following differential equation for $\Psi^{(L)}$,
\be
(1 - u - v) u v \partial_u \partial_v \Psi^{(L)}(u,v) = \Psi^{(L-1)}(u,v) \,, 
\label{Lpldiffeq}
\ee
where we define $\Psi^{(0)}(u,v) \equiv 1-u-v$.

We can also consider pentaladder integrals with a massive corner on the pentagon subintegral, see Fig. \ref{fig-pentaladders}(b). At one loop the integral takes the form
\be\label{1pentmassive}
\tilde{F}^{(1)}_{\rm pl} (x_1,x_3,x_4,x_6,x_7) = \int \frac{d^4x_r }{i \pi^2} \frac{x_{ar}^2}{x_{1r}^2 x_{3r}^2 x_{4r}^2 x_{6r}^2 x_{7r}^2}\,.
\ee
Here the point $x_a$ is light-like separated from $x_3,x_4,x_6,x_7$. It is convenient to write this integral as
\be\label{defpsitildeuvw}
\tilde{F}^{(1)}_{\rm pl}(x_1,x_3,x_4,x_6,x_7) = \frac{x_{1a}^2}{x_{14}^2 x_{16}^2 x_{37}^2} \frac{\tilde{\Psi}^{(1)} (u,v,w)}{(1-u-v+uvw)}\, ,
\ee
where the three cross-ratios are
\be
u=\frac{x_{13}^2 x_{47}^2}{x_{14}^2 x_{37}^2} \, , \quad v= \frac{x_{17}^2 x_{36}^2}{x_{16}^2 x_{37}^2} \, , \quad w=\frac{x_{37}^2 x_{46}^2}{x_{36}^2 x_{47}^2}\,.
\ee
We remark that the function $\Psi^{(1)}(u,v)$ discussed before can be obtained from $\tilde{\Psi}^{(1)}(u,v,w)$ by taking the (smooth) soft limit $w \to 0$.

Following the same logic as before we obtain the equation
\be
u \partial_u v \partial_v \tilde{\Psi}^{(1)}(u,v,w) = 1\,.
\ee
A similar analysis holds for the $L$-loop case defined by
\be\label{looppentmassive}
\tilde{F}^{(L)}_{\rm pl}(x_1,x_3,x_4,x_6,x_7) = \int \frac{d^4x_r}{i \pi^2} \frac{ x_{ar}^2 x_{37}^2}{x_{3r}^2 x_{4r}^2 x_{6r}^2 x_{7r}^2} \tilde{F}^{(L-1)}(x_1,x_3,x_r,x_7)\,.
\ee
Writing the integral as a function of the cross-ratios,
\be
\tilde{F}^{(L)}_{\rm pl}(x_1,x_3,x_4,x_6,x_7) = \frac{x_{1a}^2}{x_{14}^2 x_{16}^2 x_{37}^2} \frac{\tilde{\Psi}^{(L)} (u,v,w)}{(1-u-v+uvw)}\, ,
\ee
we find the differential equation,
\be
(1-u-v+uvw)uv\partial_u \partial_v \tilde{\Psi}^{(L)}(u,v,w) = \Psi^{(L-1)}(u,v,w)\,, 
\ee
with $\tilde{\Psi}^{(0)}(u,v,w)\equiv 1-u-v+uvw$.

\subsubsection{Pentaladders reloaded}
\label{sect-pentaladders-reloaded}

\medskip

In the case of the pentaladder integrals we can also arrive at the differential equation by considering the momentum twistor representation of the integrals.
A clue to constructing the right operator comes from the fact that the Laplacian naively annihilates the integrand,
\be
\Box_1 \frac{1}{x_{1r}^2} = 0 \quad \text{ (naive)} \,.
\ee
Of course we have seen that this does not imply that the integral itself is annihilated by the Laplace operator because there is an anomaly in the form of the delta function as in (\ref{Laplacedelta}).

Let us recall that if one writes the one-loop finite pentagon
integral using momentum twistors it takes the form,
\be \Psi^{(1)}(u,v) =F_{7;1,3,5}^{\rm pent} = \int \frac{d^4
Z_{AB}}{i \pi^2} I^{\rm pent}_{7;1,3,5}\, , \label{twistorpsi1} 
\ee
\be I_{7;1,3,5}^{\rm pent}=\frac{\fbr{456[7} \fbr{1]234}
\fbr{AB35}}{\fbr{AB71}\fbr{AB23}\fbr{AB34}\fbr{AB45}\fbr{AB56}}\,.
\label{pentintegrand} \ee It is possible to construct an operator
acting on the momentum twistor variables which also annihilates the
integrand $I^{\rm pent}_{7;1,3,5}$. Let us first introduce some
notation to deal with twistor derivatives.

We define a twistor derivative $O_{ij}$ as
\begin{equation}
O_{ij} = Z_i\cdot \frac{\partial}{\partial Z_j}\,,
\end{equation}
It acts trivially on four-brackets, $O_{ij}\fbr{j\ell k m} =
\fbr{i\ell km}$. The normalized integrals are homogenous in all
external twistors, which can be also written as
\begin{equation}
O_{ii} I = 0\,,\qquad\qquad i=1,\dots n \,.
\end{equation}
Let us do a trivial exercise which will be important in the following
discussion. Let us start with the rational function of four-bracket,
$I = \fbr{AB24}/\fbr{AB23}\fbr{AB34}$ and act on it with the
operator $O_{12}$. We immediately get
\begin{equation}
O_{12} I = \frac{\fbr{AB14}}{\fbr{AB23}\fbr{AB34}} -
\frac{\fbr{AB24}\fbr{AB13}}{\fbr{AB23}^2\fbr{AB34}} =
\frac{\fbr{AB12}}{\fbr{AB23}^2}
\end{equation}
where we used Shouten identity
$\fbr{AB14}\fbr{AB23}-\fbr{AB24}\fbr{AB13} = \fbr{AB12}\fbr{AB34}$.
Now, we see that the dependence of twistor $Z_3$ is just through the
four-bracket $\fbr{AB23}$. If we now act with the operator $O_{23}$,
then we get zero, therefore
\begin{equation}
O_{23}O_{12} \frac{\fbr{AB24}}{\fbr{AB23}\fbr{AB34}} = 0\,.
\end{equation}
Most of the equations we will derive later are based on the same
principle.

In the case of the pentagon integral the analysis above leads us to define the following operator
\be
\tilde{O}_{234} = N_{\rm pl} O_{34}O_{23} N_{\rm pl}^{-1}\, ,\qquad N_{\rm pl}=\fbr{456[7} \fbr{1]234}\,.
\ee
The operator $\tilde{O}_{234}$ naively annihilates the integrand $I^{\rm pent}_{n;1,3,5}$,
\be
\tilde{O}_{234} I^{\rm pent}_{7;1,3,5} = 0 \qquad \text{(naive)} \,.
\label{OOnaive}
\ee
We have written `naive' here because, just as for the Laplace operator, we must remember that we are going to perform an integration and there may be anomalies like the delta function left behind. In fact we can easily convince ourselves that there are indeed anomalous terms in place of the naive zero on the RHS of (\ref{OOnaive}). Let us express the action of the operator on the integral function $\Psi^{(1)}(u,v)$. We recall that the cross-ratios take the following form in terms of momentum twistors,
\be
u= \frac{\fbr{7123} \fbr{3456}}{\fbr{7134}\fbr{2356}}\,,\,\quad v =\frac{\fbr{7156}\fbr{2345}}{\fbr{7145} \fbr{2356}}\,.
\ee
Acting with the operator $\tilde{O}$ on the integral function we find
\be
\tilde{O}_{234} \Psi^{(1)}(u,v) = -\frac{\fbr{7135}N_{\rm pl}}{\fbr{7145}\fbr{7134}\fbr{3456}} u  \partial_u v \partial_v  \Psi^{(1)}(u,v)\,.
\label{expltwistorop}
\ee
We already know from acting with the Laplace operator that the RHS of (\ref{expltwistorop}) is not zero. Indeed the second-order derivative of $\Psi^{(1)}$ is $1$ according to (\ref{psi1pde}). Writing the equation without the factors of $N_{\rm pl}$ on both sides leads us to the following relation,
\begin{align}
O_{34} O_{23} \int \frac{d^4Z_{AB}}{i\pi^2} \frac{\fbr{AB35}}{\fbr{AB71}\fbr{AB23}\fbr{AB34}\fbr{AB45}\fbr{AB56}}
= -\frac{\fbr{7135}}{\fbr{7134}\fbr{7145}\fbr{3456}} \,.
\label{subintidentity}
\end{align}
This equation can then be used whenever we find the one-loop pentagon as a subintegral. In particular one can derive the differential equations for the multi-loop pentaladder integrals. As an example we discuss here the finite two-loop pentabox integral,
\begin{align}
&\Psi^{(2)}(u,v)  = \notag\\
&\int \frac{d^4 Z_{AB}}{i\pi^2}\frac{d^4 Z_{CD}}{i\pi^2} \frac{\fbr{AB35}N_{\rm pl}\fbr{2356}}{\fbr{AB23}\fbr{AB34}\fbr{AB45} \fbr{AB56} \fbr{ABCD} \fbr{CD56}\fbr{CD71}\fbr{CD23}}\,.
\end{align}
Applying the operator $\tilde{O}_{234}$ and using the identity (\ref{subintidentity}) we find
\be
\tilde{O}_{234} \Psi^{(2)}(u,v) = -\frac{\fbr{2356}}{\fbr{3456}} \int \frac{d^4Z_{CD}}{i \pi^2} \frac{N_{\rm pl} \fbr{CD35}}{\fbr{CD71}\fbr{CD23}\fbr{CD34}\fbr{CD45}\fbr{CD56}}.
\ee
Using (\ref{expltwistorop}) on the LHS and the definition of the one-loop pentagon (\ref{twistorpsi1}), (\ref{pentintegrand}) on the RHS we arrive at
\be
\frac{\fbr{7135}N_{\rm pl}}{\fbr{7134}\fbr{7145}\fbr{2356}} u  \partial_u v \partial_v  \Psi^{(2)}(u,v) = \Psi^{(1)}(u,v)\,.
\ee
By using cyclic identities one finds that the four-brackets on the LHS can be written in terms of $u$ and $v$ and we arrive at the equation
\be
(1-u-v)u\partial_u v \partial_v  \Psi^{(2)}(u,v) = \Psi^{(1)}(u,v)\,,
\ee
exactly as derived from the Laplace operator in (\ref{Lpldiffeq}) in the case $L=2$. We have rederived the equation for the two-loop finite pentaladder to simplify the equations but the derivation for $L$ loops is essentially identical and leads to (\ref{Lpldiffeq}) for general $L$.\\

In summary, we have derived the second-order equations (\ref{Lpldiffeq}) by acting with the twistor differential operator $O_{34} O_{23}$ on the
pentagon sub-integral, thanks to equation (\ref{subintidentity}). 
What this implies is that whenever we have an integral with the pentagon integral as a sub-integral, we can use this mechanism and generate
a second-order equation by using (\ref{subintidentity}). Importantly, this is also possible in cases where one cannot apply the Laplace operator,
as e.g. for the double pentagon integrals that we will discuss in section \ref{sect-double-pentaladders}.
Before doing this, we are going to discuss the generalisation where the pentagon sub-integral has a massive corner.\\

In the case of the pentaladders with a massive corner one again finds certain operators which annihilate the integrand. Let us consider the integrand of the one-loop pentagon with three massive legs,
\be
I^{\rm pent}_{8;1,3,6} = \frac{N_{\rm pm} \fbr{AB36}}{\fbr{AB81}\fbr{AB23}\fbr{AB34}\fbr{AB56}\fbr{AB67}}\,,
\ee
with $N_{\rm pm} = \fbr{567[8}\fbr{1]234}$.
The integrand is annihilated by certain twistor operators,
\be 
N_{\rm pm} O_{24} O_{42} N_{\rm pm}^{-1} I^{\rm pent}_{8;1,3,6} = 0\,,\qquad N_{\rm pm} O_{75}O_{57} N_{\rm pm}^{-1} I^{\rm pent}_{8;1,3,6} = 0\,.
\ee
Applying these operators to the explicit function we actually find that they annihilate it, i.e. there is no anomaly associated to this operator on $F^{\rm pent}_{8;1,3,6}$. In fact we will see that $F^{\rm pent}_{8;1,3,6}$ is given explicitly by \cite{refNimanew}\footnote{Here we are assuming that all dual distances are spacelike. One must carefully analytically continue to the regions where some of them become timelike.} 
\begin{eqnarray}\label{mass-pent-explicit}
F^{\rm pent}_{8;1,3,6} \equiv \tilde{\Psi}^{(1)}(u,v,w)&=&   \log u \log v + \Li_{2}(1-u ) +\Li_{2}(1-v ) +\Li_{2}(1-w )  \nonumber \\
&&- \Li_{2}(1- u w ) -\Li_{2}(1- v w )  \,,
\end{eqnarray}
where the cross-ratios take the form
\be
u=\frac{\fbr{8123} \fbr{3467}}{\fbr{8134} \fbr{2367}} \, , \quad v= \frac{\fbr{8167} \fbr{2356}}{\fbr{8156} \fbr{2367}} \, , \quad w=\frac{\fbr{2367} \fbr{3456}}{\fbr{2356} \fbr{3467}}\,.
\ee
We remark that the function $ \Psi^{(1)}(u,v)$ describing the seven-point integral $F_{7;1,3,5}^{\rm pent}$ can be obtained
from $\tilde{\Psi}^{(1)}(u,v,w)$ by taking a (smooth) soft limit,
\be
\lim_{w \to 0}  \tilde{\Psi}^{(1)}(u,v,w) =  \Psi^{(1)}(u,v)\,.
\ee
Using (\ref{mass-pent-explicit}) we then find that the following homogeneous equations are satisfied,
\be \label{eq-mass-pent1}
N_{\rm pm} O_{24} O_{42} N_{\rm pm}^{-1} F^{\rm pent}_{8;1,3,6} = 0\,, \qquad N_{\rm pm} O_{75} O_{57} N_{\rm pm}^{-1} F^{\rm pent}_{8;1,3,6} = 0\,.
\ee
Writing these in terms of the cross-ratios we find,
\begin{align}
\bigl[& (1-u)^2 \partial_u u \partial_u +  (1-w)^2 \partial_w w \partial_w - uv(1-u)(1-w) \partial_u \partial_v \notag \\
&+ (1-u)(1-w)(1+uw) \partial_u \partial_w - v(1-w)^2 \partial_v \partial_w\bigr] \tilde{\Psi}^{(1)}(u,v,w) = 0 \,,
\end{align}
and the equation obtained from swapping $u$ and $v$.
As before, we can generalise these equations to higher-loop integrals.
Note, however, that e.g. the first operator given in (\ref{eq-mass-pent1}) , $O_{24} O_{42}$, contains a derivative
acting on $Z_{2}$. We can only use (\ref{eq-mass-pent1}) on a pentagon sub-integral of a higher-loop integral
if the latter does not depend on $Z_{2}$ elsewhere, to avoid cross terms. This is the case if there are at least two additional
legs attached at the intersection of the pentagon integral with the rest of the diagram (both additional legs must be on the same side of
the diagram).\\

We will now present a further mechanism for generating differential equations that will yield an equation 
that is valid without the above restriction.
To begin with, let us apply the differential operator $O_{24} O_{75}$ to the integrand $I^{\rm pent}_{8;1,3,6}$. A short 
calculation shows that
\begin{equation}
O_{24} O_{75} I^{\rm pent}_{8;1,3,6} =  N_{\rm pm}\,  \frac{ \fbr{AB36} }{\fbr{AB81} \fbr{AB34}^2 \fbr{AB56}^2 }\,,
\end{equation}
i.e. we have cancelled two propagators.
Notice that for obtaining this expression no cyclic identities were needed. What this means is that 
the calculation would go through unchanged in the presence of a $+m^2$ regularisation, so that
we do not expect contact terms here.
Also note that the resulting integral is finite thanks to the numerator factor. 
Since it is also dual conformal,
it can only have a very restricted variable dependence. In fact, one can easily convince oneself
that up to a trivial factor, this integral is a dual conformal three-point function that can depend
on the dual points $x_{1}, x_{4}, x_{6}$ only. Since there are no cross-ratios at three points,
it must be a rational function.
Let us compute the latter. Introducing Feynman parameters, we find
\begin{eqnarray}
  \int d\alpha_{i} \frac{ \alpha_{1}  \alpha_{4} \alpha_{6} \fbr{3618} }{[ \alpha_{1} \alpha_{4} \fbr{8134} + \alpha_{1} \alpha_{6} \fbr{8156} + \alpha_{4} \alpha_{6} \fbr{3456} ]^3}\,.
\end{eqnarray}
The remaining projective integrals are easily carried out, with the result being proportional to
\begin{eqnarray}
 \frac{\fbr{3618}}{\fbr{8134} \fbr{8156}\fbr{3456} }\,.
\end{eqnarray}
Since the differential operator commutes with the normalisation of the integrand we find
\be\label{de-pent-massive-twistor}
O_{24} O_{75} F^{\rm pent}_{8;1,3,6}  \propto N_{\rm pm} \frac{\fbr{3681}}{\fbr{8134} \fbr{8156}\fbr{3456} }\,.
\ee
One can easily check this equation on the explicit function (\ref{mass-pent-explicit}).

The equation for the sub-integral can immediately be applied to higher-loop integrals
containing it. There is no subtlety since the twistor derivatives only act on $Z_{4}$ and $Z_{5}$, and the latter 
variables are separated from the rest of the integral.\\

We can write down several other operators which follow the same pattern. The idea is simple:
if we can manage to reduce the number of propagators to three, dual conformal symmetry
will imply that the resulting integral is rational.\footnote{Naively one might find it surprising that dual
conformal symmetry is helping us here,
since the momentum twistors already
made that symmetry manifest. Indeed with six twistor variables it is possible to form homogeneous 
cross-ratios but the final result contains no extra functional dependence on these variables.
The reason is that we are combining the symmetry with the fact that our functions
are Feynman integrals built from propagators, which leads to additional restrictions on the possible dependence on the twistor variables.} 
An example is given by $O_{42} O_{75}$. Note, however, that in that case the corresponding equation can
only be used iteratively in cases where there is no $Z_{2}$ dependence in the remainder of the integral.\\

In conclusion, whenever we encounter a pentagon sub-integral of the type considered in this section,
either with or without massive corner, we can always find at least one differential equation that can be
used iteratively. In some cases we can even find further equations.
In particular what this means is that we can find differential equations for all double pentagon integrals 
appearing in the MHV amplitudes. In the next section we are going to spell these equations out explicitly
for the case of an infinite class of double pentaladder integrals.

\subsubsection{Double pentaladders}
\label{sect-double-pentaladders}

 \begin{figure}[t]
\psfrag{dots}[cc][cc]{$\ldots$}
 \psfrag{x1}[cc][cc]{$x_1$}
 \psfrag{x2}[cc][cc]{$x_2$}
 \psfrag{x3}[cc][cc]{$x_3$}
 \psfrag{x4}[cc][cc]{$x_4$}
 \psfrag{x5}[cc][cc]{$x_5$}
 \psfrag{x6}[cc][cc]{$x_6$} \centerline{
 { \epsfysize3cm \epsfbox{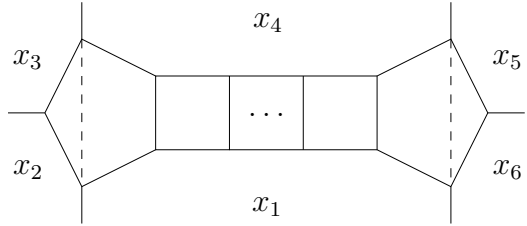} }
} \caption{\small Six-point double pentaladder integral (\ref{def-6pt-dbl-pentaladder}).} \label{fig-double-pentaladders}
\end{figure}

The fact that we have shown that one can derive differential equations using the momentum twistors is important because it allows us to find equations in cases where no Laplace operator derivation exists. As an example we will consider the six-point finite double pentaladder integrals. These integrals can be defined in terms of the pentaladders of the previous subsection,
\be \label{def-6pt-dbl-pentaladder}
\tilde{F}^{(L)}_{\rm dp}(x_1,x_2,x_3,x_4,x_5,x_6) = \int \frac{d^4x_r x_{ar}^2 x_{14}^2}{x_{1r}^2x_{2r}^2 x_{3r}^2 x_{4r}^2} \tilde{F}_{\rm pl}^{(L-1)}(x_r,x_4,x_5,x_6,x_1)\,,
\ee
see Fig. \ref{fig-double-pentaladders}.
Conformal symmetry in the $x$ variables tells us that we can write
\be
\tilde{F}^{(L)}_{\rm dp}(x_1,x_2,x_3,x_4,x_5,x_6) = \frac{x_{ab}^2}{x_{26}^2 x_{35}^2 x_{14}^2} \Omega^{(L)}(u,v,w)\,,
\ee
where $u,v$ and $w$ are the three cross-ratios,
\be
u=\frac{x_{13}^2 x_{46}^2}{x_{14}^2 x_{36}^2}\, ,\quad v = \frac{x_{24}^2 x_{15}^2}{x_{14}^2 x_{25}^2}\, , \quad w=\frac{x_{35}^2x_{26}^2}{x_{25}^2 x_{36}^2}\,.
\ee
To derive differential equations for these integrals we need to express them in momentum twistor notation. We will consider the finite six-point two-loop double pentagon integral as an example,
\begin{align}
\label{6ptdblpent}
&\Omega^{(2)}(u,v,w) = F_{6;1,3,4,6}^{\rm double-pent} \\
&= \int \frac{d^4Z_{AB}}{i\pi^2} \frac{d^4 Z_{CD}}{i\pi^2} \frac{\fbr{AB13}\fbr{CD46} N_{\rm dp}}{\fbr{AB61}\fbr{AB12}\fbr{AB23}\fbr{AB{34}} \fbr{ABCD} \fbr{CD45} \fbr{CD56} \fbr{CD61}}\,, \notag
\end{align}
where $N_{\rm dp}=\fbr{2345}\fbr{5612}\fbr{3461}$. In terms of the momentum twistors the three cross-ratios take the form,
\be
u=\frac{\fbr{6123}\fbr{3456}}{\fbr{6134}\fbr{2356}}\, , \quad v= \frac{\fbr{1234}\fbr{4561}}{\fbr{1245}\fbr{6134}}\, , \quad w=\frac{\fbr{2345}\fbr{1256}}{\fbr{2356}\fbr{1245}}\,.
\ee
We see that the double pentagon integral contains the one-loop pentagon that we studied in the previous subsection as a subintegral. Thus we can apply the same kind of differential operator and derive a differential equation. The operator we wish to apply here is
\be
\tilde{O}_{612} = N_{\rm dp} O_{12} O_{61} N_{\rm dp}^{-1}\,.
\ee
Acting with this operator on the the double pentagon integral and using the relation (\ref{subintidentity}) with the appropriate relabelling we find
\begin{align}
&\tilde{O}_{612} \Omega^{(2)}(u,v,w) \notag \\
&= -\frac{1}{\fbr{1234}}\int \frac{d^4 Z_{CD}}{i\pi^2} \frac{N_{\rm dp}\fbr{CD13}\fbr{CD46}}{\fbr{CD12} \fbr{CD23}\fbr{CD34}\fbr{CD45}\fbr{CD56}\fbr{CD61}} \nonumber\\
&= -\frac{\fbr{3461}}{\fbr{1234}} \Omega^{(1)}(u,v,w)\,.
\label{6ptdblpenteq}
\end{align}
Here $\Omega^{(1)}(u,v,w)$ is the one-loop finite hexagon integral,
\be
\Omega^{(1)}(u,v,w) = \int \frac{d^4 Z_{CD}}{i\pi^2} \frac{N_{\rm hex} \fbr{CD13}\fbr{CD46}}{\fbr{CD12}\fbr{CD23}\fbr{CD34}\fbr{CD45}\fbr{CD56}\fbr{CD61}}\,,
\ee
where $N_{\rm hex} = \fbr{2345}\fbr{5612}$.\\

{}From the discussion in section \ref{sect-pentaladders-reloaded} it is clear that
the above equations can be straightforwardly generalised to the case where the pentagon sub-integrals
have a massive middle leg.










\subsection{Divergent integrals}
\label{sect-diff-div}

Some of the integrals appearing in the MHV amplitudes are infrared divergent.
These divergences are regulated by masses as explained in section \ref{sect-moti}.
In this section we show that one can find similar differential equations as above,
that hold up to $\cO(m^2)$.

\subsubsection{One-loop box integrals}
\label{sect-boxes-oneloop}

As explained in section \ref{sect-moti}, they are special cases of the general
pentagon topology with magic numerator, when no
external legs are attached to the rightmost corner.
In this case, one obtains a  two-mass hard box integral, whose integrand
is given by
\begin{equation}\label{2mh}
F_{n;i,2}^{\rm box} = \int \frac{d^{4}x_{0}}{i \pi^2}\, \frac{x_{24}^2 x_{3i}^2}{(x_{20}^2 +m^2) (x_{30}^2 +m^2)  (x_{40}^2 + m^2)  (x^2_{i0} + m^2)}\,,
\end{equation}
Unlike the generic pentagon integral $F_{n;i,j,k}^{\rm pent}$ the two-mass hard integral is infrared divergent
and depend explicitly on the mass $m$ that regulates the infrared divergences,
see section \ref{sect-moti}.
Here we discuss that the differential equations we found for the finite integrals
work in exactly the same way here, up to $\cO(m^2)$ corrections.
To understand why this is the case, consider again the Laplace operator, but acting on a massive propagator.\footnote{In this case one could also
note that massive propagators are formally identical to free AdS bulk-to-boundary propagators, and use the corresponding
equation of motion, see \cite{D'Hoker:2002aw}.
For this it is necessary to consider situation with generic masses $m_{i}$, but this is no restriction thanks to
dual conformal symmetry. This would yield equations that are exact in the $m_{i}$, i.e. that do not rely on taking
the masses small w.r.t. the Mandelstam invariants. It is an interesting open question how to extend this efficiently to the on-shell case.}
We have
\begin{eqnarray}\label{laplacemassive}
\square_{1} \frac{1}{x_{1i}^2 + m^2} &=& -8 \frac{m^2}{(x_{1i}^2+m^2)^3 }   = - 4 \, i \pi^2  \, \delta^{(4)}(x^{\mu}_{1i}) + \cO(m^2)\,,
\end{eqnarray}
so that in this case we would indeed ``undo'' one integration as in the massless case, but up to $\cO(m^2)$ corrections.
As we discuss presently, this leads
second-order differential
equations for this integral, at least for $n\ge 5$.
We will first discuss the generic case where (\ref{2mh})
represents indeed a two-mass hard integral, and
subsequently discuss the degenerate one-mass case
where $i=4$ or $i=n$.

In the former case, there is one off-shell leg that we can apply
the Laplace operator to, see equation (\ref{laplacemassive}).
This
leads to a differential equation valid to $\cO(m^2)$, namely
\begin{equation}\label{2mhlaplace}
 x^2_{2i} x^2_{3i} x^2_{4i}  \, \square_{i} \, (x_{24}^2 x_{3i}^2)^{-1}\, F_{n;i;2}^{\rm box}  = -4  + \cO(m^2)\,.
\end{equation}
We can explicitly verify (\ref{2mhlaplace}) by differentiating the known
answer for the two-mass hard integral.
In terms of Feynman parameters, it is given by
\begin{eqnarray}
F_{n;i;2}^{\rm box}  =  \int d\alpha_{i} \frac{x_{24}^2 x_{3i}^2 \, \delta(\alpha_2 + \alpha_3 + \alpha_4 + \alpha_i -1) }{\lbrack \alpha_2 \alpha_4 x_{24}^2
+ \alpha_i ( \alpha_2 x_{2i,m}^2 + \alpha_3 x_{3i,m}^2 + \alpha_4 x_{4i,m}^2 ) + m^2 \rbrack^2}\,,
\end{eqnarray}
where $x_{2i,m}^2 := x_{2i}^2 + m^2$, etc.
One obtains
\begin{equation}\label{2mh-solution}
F_{n;i;2}^{\rm box}  =
\frac{1}{2} \log^2 \left( { y_1 y_2 y_3} \right)
+  2\, {\rm Li}_{2}\left(1 - y_2 \right)
+  2\, {\rm Li}_{2}\left(1 - y_3 \right) + \cO(m^2)\,,
\end{equation}
where we used the variables $y_1 =   x_{24}^2 / m^2 \,, y_2 = x_{3i}^2/ x_{2i}^2 \,, y_3 = x_{3i}^2/ x_{4i}^2$.
Just like in section \ref{sect-diff-finite}, we can also find twistor derivatives that lead to differential equations.
However, the two-mass hard integral is infrared divergent, and here we have dropped certain $\cO(m^2)$ terms.
When applying differential operators to an infrared divergent subintegral, e.g. in the double box case, 
one must be very careful when making such approximations. We postpone a detailed analysis of the
double box integrals to future work.

\subsubsection{Penta-box integrals}
\label{sec-divpentabox}

 \begin{figure}[t]
\psfrag{dots}[cc][cc]{$\ldots$}
 \psfrag{x1}[cc][cc]{$x_1$}
 \psfrag{x2}[cc][cc]{$x_2$}
 \psfrag{x3}[cc][cc]{$x_3$}
 \psfrag{x4}[cc][cc]{$x_4$}
 \psfrag{x5}[cc][cc]{$x_5$}
 \centerline{
 { \epsfysize3cm \epsfbox{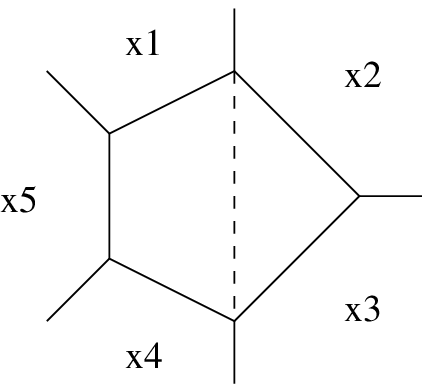} \hspace{2cm} \epsfysize3cm \epsfbox{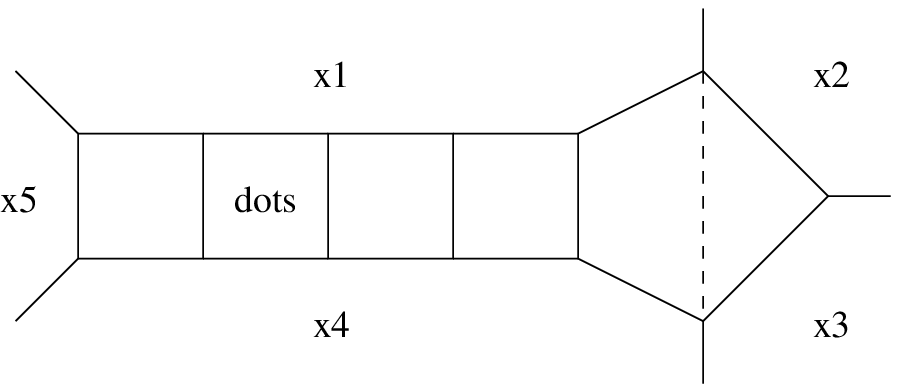}}
} \caption{\small Five-point penta-ladder integrals considered in the text.} \label{fig-5pt-pentaladders}
\end{figure}

A class of finite pentaladder integrals was already discussed in section \ref{sect-diff-finite}.
Here we show that the IR divergent penta-box integrals also satisfy analogous differential
equations, up to $\cO(m^2)$.

Let us consider the pentaladder integrals shown in Fig. \ref{fig-5pt-pentaladders}. The discussion
for other divergent pentaladder integrals would be very similar.
At one loop, we have
\begin{equation}
F^{\rm pent}_{5;5,1,3} = \int \frac{d^{4}Z_{AB}}{i \pi^2} \, \frac{ \fbr{2345} \fbr{4512} \fbr{AB13} }{\fbr{51AB} \fbr{12AB} \fbr{23AB} \fbr{34AB} \fbr{45AB} } \,,
\end{equation}
Since all external legs are on-shell, we cannot use the Laplace operator.
We already know from section \ref{sect-pentaladders-reloaded} that in
such cases twistor derivatives can have the same effect. Indeed, consider
the action of the differential operator
\begin{eqnarray}\label{o5112}
N O_{32} O_{43}  N^{-1} =  Z_{3} \cdot \frac{\partial}{ \partial Z_{2}} \; Z_{4} \cdot \frac{\partial}{\partial Z_{3}} + \frac{\fbr{ 5134 }}{\fbr{ 1245 } } Z_{4} \cdot \frac{\partial}{\partial Z_{3}} \,,
\end{eqnarray}
where $N= \fbr{2345} \fbr{4512}$,
One can check that it
annihilates its integrand, up to $\cO(m^2)$ terms.
As discussed above, one can think of the mass as regulating distributional terms that
arise when acting with the derivatives.
Let us now find out what happens when we act with the operator in (\ref{o5112})
on the function obtained after integration.
Computing the one-loop integral $F^{\rm pent}_{5;5,1,3}$ directly
using Feynman parameters in the small $m^2$ limit, we have (see Appendix \ref{app-5pt})
\begin{equation}\label{magic-5ptpentagon-nice-main}
F^{\rm pent}_{5;5,1,3}  =  -\frac{1}{2}  \log^2(y_1 y_2 y_3) -2 \, {\rm Li}_{2}( 1 - y_1) -2 \, {\rm Li}_{2}( 1 - y_2) + \frac{\pi^2}{6}  +\cO(m^2) \,.
\end{equation}
Here $y_1=x^2_{35}/x^2_{13}, y_2 = x^2_{25}/x_{24}^2 , y_{3}= x^2_{14}/m^2$, and we recall the relation of
the dual variables to the momentum twistors,
\begin{equation}
x_{13}^2 = \frac{\fbr{ 5123 } }{\tbr{ 51} \tbr{ 23 }} \,,x_{24}^2 = \frac{ \fbr{ 1234} }{ \tbr{12} \tbr{ 34}} \,, x_{35}^2 = \frac{ \fbr{ 2345 } }{ \tbr{ 23} \tbr{ 45}} \,,x_{14}^2 = \frac{ \fbr{3451}}{ \tbr{ 34}\tbr{ 51}}\,,x_{25}^2 = \frac{ \fbr{4512}}{\tbr{45}\tbr{ 12}} \,.
\end{equation}
Using the chain rule and acting on (\ref{magic-5ptpentagon-nice-main}) one can easily verify that
\begin{eqnarray}\label{divpent1loop}
\frac{\fbr{5123}}{\fbr{4513}} \left[ Z_{3} \cdot \frac{\partial}{ \partial Z_{2}} \; Z_{4} \cdot \frac{\partial}{\partial Z_{3}} + \frac{\fbr{ 5134 }}{\fbr{ 1245 } } Z_{4} \cdot \frac{\partial}{\partial Z_{3}}  \right] \, F^{\rm pent}_{5;5,1,3}  =  -1
+\cO(m^2)\,,
\end{eqnarray}
i.e. the r.h.s. is not zero but a simple rational term. In other words, it is still true that
the differential operator ``undoes'' the integration, at least up to $\cO(m^2)$.\\

We now discuss the all-loop generalisation of these equations.
The two-loop version of $F^{\rm pent}_{5;5,1,3}$ is
\begin{eqnarray}
F^{\rm penta-box}_{5;5,1,3} \hspace{-0.2cm} &=& \hspace{-0.2cm}  \int  \frac{d^{4}Z_{AB} d^{4}Z_{CD} \, (i \pi^2)^{-2}\, N \fbr{CD13}  }{ \fbr{AB34} \fbr{AB45} \fbr{AB51} \fbr{ABCD} \fbr{CD51}\fbr{CD12}\fbr{CD23}\fbr{CD34}} \,,
\end{eqnarray}
where $N=\fbr{3451} \fbr{2345} \fbr{4512}$.

As in the previous sections,
we need to discuss the pentagon subintegral.
In fact the subintegral is the same finite integral that was discussed in section \ref{sect-diff-finite}, with
the difference that since the two-loop integral is IR divergent its propagators contain $+m^2$ pieces.
However, we already know that the special numerator makes the subintegral finite, and therefore
the effect of the masses in the pentagon propagator should be only $\cO(m^2)$. \footnote{This argument presupposes
that the divergent box subintegral produces only logarithmic but no linear divergences in $m^2$.
This is the case for all integrals we consider.} Therefore
we can neglect these masses when acting with the differential operator, and it is clear
that we will obtain differential equations completely analogous to those discussed in section \ref{sect-diff-finite}.
The only difference is that by construction they are only valid up to $\cO(m^2)$ corrections.
Therefore we have
\begin{equation}\label{divpent2loop}
\frac{\fbr{5123}}{\fbr{4513}} N O_{32} O_{43} N^{-1} \, F^{\rm penta-box}_{5;5,1,3}  =
F^{\rm pent}_{5;5,1,3}  +\cO(m^2)\,.
\end{equation}
Of course one can immediately generalise this equation to arbitrary loop order,
by replacing the integral on the LHS and RHS by the $L$-loop and $(L-1)$-loop penta-box integrals,
respectively.
Finally, the other divergent penta-box integrals which differ in the arrangement of external legs
all satisfy analogous differential equations.\\

In summary, we have found differential equations for all integrals contributing to
the one- and two-loop MHV amplitudes in $\cN=4$ SYM, except for the
six double box integrals.
The differential equations we find are iterative in the loop order and apply to
infinite series of on-shell loop integrals. We expect that integrals contributing to
higher-loop MHV amplitudes and other helicity configurations are constrained in
a similar way.

The same mechanisms for generating differential equations that we described
above should also hold for the double box integrals.
However a separate analysis is
required there since they do not contain IR finite subintegrals. We leave this for 
future work.
(An encouraging fact is that in Appendix \ref{sect-diff-integrands} we find 
differential operators that naively annihilate the double box integrand.)
Of course, the double box integrals are in some sense
the simplest integrals appearing in the two-loop MHV amplitudes, so perhaps
they could also be computed by other means. Also, they are related to the
more general integrals that do satisfy differential equations through soft limits.
If one can control the non-commutativity of the soft limit with the regulator
limit this can also provide a useful tool for computing them.\\

\section{Solving the differential equations}
\label{sect-solve}

\subsection{Harmonic Polylogarithms}
We will now discuss solving the differential equations. The solutions we find are typically written in terms of polylogarimic functions. Here we will give  a brief introduction to the notation we use which can be found in \cite{Remiddi:1999ew,Gehrmann:2000zt}. 

The simple harmonic polylogarithm functions are given in terms of iterated integrals. We start from the functions
\be
H(0;x)=\log x\,, \qquad H(1;x)= -\log (1-x)\,.
\ee
Their derivatives are given by the fractions
\be
\frac{d}{dx} H(a;x) = f(a;x)\,,\qquad f(0;x)=\frac{1}{x}\,, \quad f(1;x) = \frac{1}{1-x}\,.
\ee
We can now define weight $w$ harmonic polylogarithms depending on a vector with entries of ones and zeros. For a vector of all zeros of length $w$ (written $\vec{0}_w$) we have
\be
H(\vec{0}_{w};x) = \frac{1}{w!}\log^w x\,.
\ee
Further harmonic polylogarithms depending on a more general vector $\vec{m}_w = (a,\vec{m}_{w-1})$ can be defined by repeated integration,
\be
H(\vec{m}_w;x) = \int_0^x dx' f(a;x') H(\vec{m}_{w-1};x')\,.
\ee
Below we save space by writing the vector $\vec{m}_w$ as a subscript.
Finally we can use the shorthand notation where proceeding from right to left a zero is eliminated if it is to the left of a non-zero entry while at the same time one is added to the value of that entry. For example
\be
H_{3,2}(x) = H(0,0,1,0,1;x)\,.
\ee
There also exist harmonic polylogarithms with negative arguments but we will not need them here.

The simple harmonic polylogarithms defined above are all single variable functions. However they can be generalised to functions of more than one variable. In general the relevant functions are the Goncharov polylogarithms. Here we will only need functions of two variables at most which can be described in terms of two-dimensional harmonic polylogarithms. These functions are defined in much the same way as the simple harmonic polylogarithms but we allow slightly more general arguments of the vector $\vec{m}_w$. This is done by enlarging the set of weight one functions and fractions to include
\be
H(1-y;x) = - \log \Bigl(1-\frac{x}{1-y}\Bigr)\,,\qquad f(1-y;x)=\frac{1}{1-x-y}\,.
\ee

\subsection{Solution for seven-point pentagon and penta-box integral}

Let us consider the equation for $\Psi^{(1)}$, the one-loop pentagon integral. The equation is
\be
uv\partial_u \partial_v \Psi^{(1)}(u,v) = 1\,.
\ee
The general solution is
\be
\Psi^{(1)}(u,v) = \log u \log v + g(u) + g(v)
\ee
for some single variable function $g$.

Let us recall that we chose to write the pentaladder integrals is the following way,
\be
\tilde{F}^{(L)}_{\rm pl}(x_1,x_3,x_{4},x_{5},x_{6}) = \frac{x_{1a}^2}{x_{14}^2 x_{15}^2 x_{36}^2} \frac{\Psi^{(L)}(u,v)}{(1-u-v)} \,.
\ee
This form makes it clear that the function $\Psi^{(L)}(u,v)$ should have a simple zero when $1-u-v=0$ since the integral clearly has no special singular behaviour on this region. Thus we have
\be
\Psi^{(L)}(u,1-u)=0\,.
\label{psibc}
\ee
The boundary condition (\ref{psibc})
implies that the function $g$ should satisfy
\be
g(u) + g(1-u) = - \log u  \log (1-u)\,.
\label{functionaleq}
\ee
The solution of (\ref{functionaleq}) is ambiguous up to any function odd in swapping $u$ and $1-u$.
Thus we find
\be
g(u) = {\rm Li}_2(1-u) - \tfrac{1}{2}\zeta(2) + h(u)-h(1-u)\,,
\ee
for any single variable function $h$.

We need some extra data to fix the function $h(u)-h(1-u)$. If we adopt maximal transcendentality (which translates here to the fact that $h(u)$ can only be a sum of harmonic polylogarithms) and also the requirement that there are no branch points of $\Psi^{(1)}(u,v)$ for $u>0,v>0$ then we fix $h(u)=0$.
Indeed if we assume that we can write $h(u)$ as a sum of harmonic polylogarithms of degree 2,
\be
h(u) = a\, {\rm Li}_2(u) + b \, \log^2 u + c \,\log u \,{\rm Li}_1(u) + d \, {\rm Li}_1^2(u)\,,
\ee
then the only solution where $h(u)-h(1-u)$ has no branch points at $u=1$ is indeed $h=0$.
Thus we deduce that
\be
\Psi^{(1)}(u,v) = \log u \log v + \Li_2(1-u) + \Li_2(1-v) - \zeta(2)\,.
\ee
Given the correct answer for $\Psi^{(1)}$ we can go on to determine the next pentaladder function $\Psi^{(2)}$. The equation we have to solve is
\be
(1-u-v)uv\partial_u \partial_v \Psi^{(2)}(u,v) = \Psi^{(1)}(u,v)\,.
\label{twolooppl}
\ee
It is helpful to divide by the factor $(1-u-v)uv$ and use partial fractions to write
\be
\partial_u \partial_v \Psi^{(2)}(u,v) = \Psi^{(1)}(u,v)\biggl[\frac{1}{uv} + \frac{1}{u(1-u-v)} + \frac{1}{v(1-u-v)}\biggr]\,.
\ee
Now we look for particular solutions which reproduce each of the three terms on the RHS, i.e.
\begin{align}
\partial_u \partial_v H^0(u,v) &= \frac{\Psi^{(1)}(u,v)}{uv}\,,\\
\partial_u \partial_v H^u(u,v) &= \frac{\Psi^{(1)}(u,v)}{u(1-u-v)}\, ,\\
\partial_u \partial_v H^v(u,v) &= \frac{\Psi^{(1)}(u,v)}{v(1-u-v)}\,.
\end{align}
We find the following solution for $H^0$,
\begin{align}
H^0(u,v) = & 2 \zeta(4) + \tfrac{1}{2}\zeta(2)^2 - \zeta(2) \bigl(\Li_2(1-u) + \Li_2(1-v) \bigr) \notag \\
&+\log u \log v \bigl( \tfrac{1}{4}\log u \log v +\Li_2(1-u) + \Li_2(1-v) - \zeta(2) \bigr) \notag\\
& + \log u \bigl( 2 H_{2,1}(1-v) - \Li_3(1-u) + \zeta(3)\bigr) \notag \\
&+ \log v \bigl( 2 H_{2,1}(1-u) - \Li_3(1-v) + \zeta(3) \bigr) \notag \\
&- 3 \bigl(H_{3,1}(1-u) + H_{3,1}(1-v) \bigr) \,,
\label{H0}
\end{align}
where we have used the notation of harmonic polylogarithms \cite{Remiddi:1999ew}. In fact $H_{2,1}$ is a Nielsen polylogarithm and can be reexpressed in terms of ordinary trilogarithms but we prefer not to do so here. For $H^u$ we find\footnote{We thank L.~Dixon for pointing out a typo in a previous version of this formula.},
\begin{align}
H^u(u,v) = &\zeta(2)\bigl(H_{0,1-v}(u) -\Li_2(u) - \log u \log(1-v)\bigr) +H_{0,0}(u) H_{1,0}(v)  \notag \\
&+ \log v \bigl( H_{0,1-v,0}(u) - H_{2,0}(u)\bigr)  + \log u H_{1,1,0}(v) \notag \\
&- H_{2,1,0}(u) + H_{0,1-v,1,0}(u)+H_{0,1-v}(u) H_{1,0}(v)
\,,
\label{Hu}
\end{align}
where in addition to harmonic polylogarithms we have used the two-dimensional harmonic polylogarithms of \cite{Gehrmann:2000zt}. We have neglected some obvious simplifications such as $H_{0,0}(u) = \tfrac{1}{2} \log^2 u$ and $H_{1,0}(u) = - \Li_2(u) - \log u \log(1-u)$. By symmetry we have also
\be
H^v(u,v) = H^u(v,u)\,.
\label{Hv}
\ee
Thus the general solution to the differential equation (\ref{twolooppl}) is
\be
\Psi^{(2)}(u,v) = H^0(u,v) + H^u(u,v) + H^v(u,v) + g(u) + g(v)\,,
\label{gensolnpsi2}
\ee
for some single variable function $g$. The function $g$ can be constrained by the fact that $\Psi^{(2)}(u,v)$ should vanish when $1-u-v=0$. Again assuming that it is purely transcendental and has no spurious branch cuts (although as presented, the branch cut structure of $H^u$ is not obvious) fixes the function. We find
\begin{align}\label{g}
g(u) = &-\tfrac{1}{2}\zeta(2)^2 - 3 \zeta(4) +\zeta(2) \bigl(\tfrac{1}{2}\log^2 u + \Li_2(1-u)\bigr) \\
&+ \log u \bigl(\Li_3(1-u) + H_{2,1}(1-u) - 3 \zeta(3)\bigr) + 2 H_{3,1}(1-u) + 3 H_{2,1,1}(1-u) \,.
\nonumber
\end{align}
We have checked that this combination of functions numerically matches the Mellin-Barnes representation of the integral in the triangular region $u>0$, $v>0$, $u+v<1$ to a high degree of accuracy. Thus we are confident that (\ref{gensolnpsi2}) together with (\ref{H0}),(\ref{Hu}),(\ref{Hv}), and (\ref{g}) correctly give the two-loop pentaladder function.

\subsection{Equations for eight-point pentagon and pentaladder integrals}

We would now like to discuss the equations for integrals with pentagon subintegrals with massive corners. For example we have the pentaladder integrals with a massive corner,
\be
\tilde{F}^{(L)}_{\rm pl}(x_1,x_3,x_4,x_6,x_7) = \frac{x_{1a}^2}{x_{14}^2 x_{16}^2 x_{37}^2} \frac{\tilde{\Psi}^{(L)} (u,v,w)}{(1-u-v+uvw)}\, ,
\ee
Note that the way we have chosen to write the RHS means that $\tilde{\Psi}^{(L)}(u,v,w)$ vanishes when $1-u-v+uvw=0$ because the integral has no pole in this region.

As we have seen for the pentaladders with a massive corner we have two sources of differential equations. Firstly from applying the Laplace operator we have:
\be
(1-u-v+uvw)uv\partial_u \partial_v \tilde{\Psi}^{(L)}(u,v,w) = \Psi^{(L-1)}(u,v,w)\,.
\label{Laptildepsi}
\ee
Then in the twistor language we can apply the operator
\be
O_{24}O_{75}
\ee
which also produces a rational function. In fact we have
\be
N_{\rm pm} O_{24} O_{75} N_{\rm pm}^{-1} \tilde{\Psi}^{(1)}(u,v,w)= O_{24}O_{75} \tilde{\Psi}^{(1)}(u,v,w) = \frac{N_{\rm pm} \fbr{3681}}{\fbr{3456}\fbr{3481}\fbr{5681}}\,.
\ee
This leads to the equation
\begin{align}
&w[(1-w)\partial_w((1-w)\partial_w + v(1-v)\partial_v + u(1-u)\partial_u) + u(1-u)v(1-v)\partial_u\partial_v]\tilde{\Psi}^{(1)}(u,v,w) \notag \\
&= 1-u-v+uvw \equiv \tilde{\Psi}^{(0)}(u,v,w)\,.
\end{align}
More generally we have for the $L$-loop pentaladder with a massive corner,
\begin{align}
&w[(1-w)\partial_w((1-w)\partial_w + v(1-v)\partial_v + u(1-u)\partial_u) + u(1-u)v(1-v)\partial_u\partial_v]\tilde{\Psi}^{(L)}(u,v,w) \notag \\
&= \tilde{\Psi}^{(L-1)}(u,v,w)\,.
\end{align}
From (\ref{Laptildepsi}) we know the result of acting with the final term in the operator on the LHS so we can rewrite the equation in a simpler way. We find
\be
w\partial_w[(1-w)\partial_w + u(1-u)\partial_u + v(1-v)\partial_v] \tilde{\Psi}^{(1)}(u,v,w) = 1-u-v\,,
\ee
or more generally for $L$ loops,
\be
w\partial_w[(1-w)\partial_w + u(1-u)\partial_u + v(1-v)\partial_v] \tilde{\Psi}^{(L)}(u,v,w) = \frac{(1-u-v)\tilde{\Psi}^{(L-1)}(u,v,w)}{1-u-v+uvw} \,.
\label{neweq}
\ee
Now let us consider (\ref{Laptildepsi}) restricted to $w=1$. We find that the operator factorises,
\be
u(1-u)\partial_u v(1-v)\partial_v \tilde\Psi^{(L)}(u,v,1) = \tilde{\Psi}^{(L-1)}(u,v,1)\,.
\ee
In the one-loop case we then have simply
\be
u\partial_u v \partial_v \tilde{\Psi}^{(1)}(u,v,1) = 1\,.
\ee
and so
\be
\tilde{\Psi}^{(1)}(u,v,1) = \log u \log v + f(u) + f(v) \,,
\ee
for some single variable function $f$. We recall that $\tilde{\Psi}^{(L)}(u,v,w)$ vanishes when $1-u-v+uvw=0$. In our case where $w=1$ this implies that $\tilde{\Psi}^{(1)}(u,v,1)$ vanishes when $u=1$ or $v=1$. Thus we have
\be
f(u) + f(1) = 0\,,
\ee
which implies that $f(u)=0$ and so
\be
\tilde{\Psi}^{(1)}(u,v,1) = \log u \log v\,.
\ee
In the two-loop case it is again simple,
\be
u(1-u)\partial_u v(1-v)\partial_v \tilde{\Psi}^{(2)}(u,v,1) = \log u \log v\,.
\ee
Integrating we find
\be
\tilde{\Psi}^{(2)}(u,v,1) = [H_{0,0}(u) + H_{1,0}(u)][H_{0,0}(v) + H_{1,0}(v)] + f(u) + f(v)\,
\ee
for some single variable function $f$. As before we have the condition that $\tilde{\Psi}^{(L)}(u,v,1)$ vanishes when $u=1$ or $v=1$. Thus we find
\be
-\zeta(2)[H_{0,0}(u)+H_{1,0}(u)] + f(u) + f(1) = 0\,.
\ee
Setting $u=1$ this tells us
\be
f(1) = -\tfrac{1}{2} \zeta(2)^2\,.
\ee
Thus we have
\be
f(u) = \zeta(2)[H_{0,0}(u) + H_{1,0}(u) + \tfrac{1}{2}\zeta(2)]
\ee
and we have solved for $\tilde{\Psi}^{(2)}(u,v,1)$ quite simply.

Now we can return to the other equation (\ref{neweq}). Let us perform the $w$ integral,
\be
[(1-w)\partial_w + u(1-u)\partial_u + v(1-v)\partial_v] \tilde{\Psi}^{(L)}(u,v,w) = X^{(L)}(u,v,w) +f_X^{(L)}(u,v)\,
\label{1stordereq}
\ee
where
\be
X^{(L)}(u,v,w) =(1-u-v) \int_1^w \frac{dt \tilde{\Psi}^{(L-1)}(u,v,t)}{t(1-u-v+uvt)}\,
\ee
and $f_X(u,v)$ is an arbitrary two-variable function. We can determine $f_X$ however by setting set $w=1$ in (\ref{1stordereq}) and we find
\be
[u(1-u)\partial_u + v(1-v)\partial_v]\tilde{\Psi}^{(L)}(u,v,1) = f^{(L)}_X(u,v)\,.
\ee
The LHS is known from the Laplace operator as shown above so given $\tilde{\Psi}^{(L)}(u,v,1)$ we know $f^{(L)}_X(u,v)$. In the one-loop case we have
\be
f^{(1)}_X(u,v) = (1-u)\log v + (1-v)\log u\,.
\ee
while
\be
X^{(1)}(u,v,w) =(1-u-v) \int_1^w \frac{dt}{t} = (1-u-v)\log w\,.
\ee
Thus (\ref{1stordereq}) becomes a first order equation for $\tilde{\Psi}^{(1)}(u,v,w)$. One can easily convince oneself that the unique solution is (\ref{mass-pent-explicit}) given the known boundary condition for $\tilde{\Psi}^{(1)}(u,v,1)$ or equivalently $\tilde{\Psi}^{(1)}(u,v,0)=\Psi^{(1)}(u,v)$.

More generally we always have a first order equation for $\tilde{\Psi}^{(L)}(u,v,w)$ and thus one can write the function in a simple integral form.

\subsection{Equations for the six-point hexagon and double-pentagon integrals}

Let us now discuss the simplest cases of the double pentaladder integrals. To start we will consider the finite six-point double-pentagon, defined in (\ref{6ptdblpent}).
In terms of the twistor variables the independent cross-ratios take the form
\be
u=\frac{\fbr{6123}\fbr{3456}}{\fbr{6134}\fbr{2356}}\, , \quad v= \frac{\fbr{1234}\fbr{4561}}{\fbr{1245}\fbr{6134}}\, , \quad w=\frac{\fbr{2345}\fbr{1256}}{\fbr{2356}\fbr{1245}}\,.
\ee
Translating the equation (\ref{6ptdblpenteq}) into an equation for the function $\Omega^{(2)}$ we find
\begin{align}
&\Bigl[\bigl(1-u-v+\frac{\tilde{\epsilon}}{2}\bigr)u v\partial_{u}\partial_{v}
+\frac{\tilde{\epsilon}-2u}{2}vw\partial_{v}\partial_{w}
+\frac{\tilde{\epsilon}-2v}{2}uw\partial_{u}\partial_{w}
-\frac{\tilde{\epsilon}w}{2}\partial_{w}(1-w)\partial_{w}\Bigr]\Omega^{(2)}(u,v,w)\notag \\
& = \Omega^{(1)}(u,v,w)\,.
\end{align}
with $\tilde{\epsilon}=\sqrt{\delta^2-4uvw}+\delta$ and 
$\delta = u+v+w-1$. 

In fact the sign of the square root is ambiguous and so the equation separates into two second order equations, one  homogeneous, the other inhomogeneous,
\begin{align}
&\Bigl[\bigl(1-u-v \bigr)u v\partial_{u}\partial_{v}
-uvw\partial_{v}\partial_{w}
-uvw\partial_{u}\partial_{w}\Bigr]\Omega^{(2)}(u,v,w) = \Omega^{(1)}(u,v,w)\,, \label{inhomOmega}\\
&\Bigl[uv\partial_u \partial_v + uw \partial_u \partial_w + vw \partial_v \partial_w - w\partial_w(1-w)\partial_w\Bigr]\Omega^{(2)}(u,v,w)=0\,.\label{homOmega}
\end{align}
In fact we can set $w=0$ in (\ref{inhomOmega}) and we find
\be
(1-u-v)uv\partial_u \partial_v \Omega^{(2)}(u,v,0) = \Omega^{(1)}(u,v,0)\,.
\ee
This equation is identical in form to that for the functions $\Psi^{(L)}(u,v)$ from the pentaladder with a massless corner. At one loop we have in fact
\be
\Omega^{(1)}(u,v,w) = \log u \log v + \Li_2(1-u) + \Li_2(1-v) + \Li_2(1-w) - 2 \zeta(2)\,,
\ee
and indeed $\Omega^{(1)}(u,v,0) = \Psi^{(1)}(u,v)$.
We have verified numerically from the Mellin-Barnes representation that the same holds at two loops, $\Omega^{(2)}(u,v,0) = \Psi^{(2)}(u,v)$, so this two-variable slice of the function $\Omega^{(2)}$ is the same function that we have solved for from the pentaladder equation.

Combining the two equations (\ref{inhomOmega},\ref{homOmega}) we can write a factorised operator,
\be
w\partial_w \Bigl[-u(1-u)\partial_u - v(1-v)\partial_v +(1-u-v)(1-w)\partial_w\Bigr]\Omega^{(2)}(u,v,w) = \Omega^{(1)}(u,v,w)\,.
\ee
After performing one integration on $w$ we have a first-order equation for the function $\Omega^{(2)}$.

\section{Conclusion and outlook}
\label{sect-outlook}

 \begin{figure}[t]
\psfrag{dots}[cc][cc]{$\ldots$}
  \centerline{
 { \epsfysize3cm \epsfbox{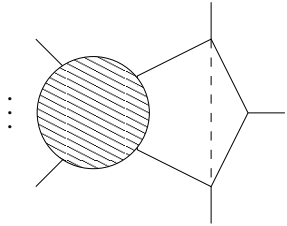} }
} \caption{\small The iterative differential equations can always be obtained if a key sub-integral is present, as shown
in the Figure. The salient feature is the presence of a ``magic numerator'', as indicated by the dashed line. 
The number of differential equations that can be obtained for an integral depends on the precise configuration
of external legs.
 } \label{fig-key}
\end{figure}

In this paper we derived a new type of differential equations for on-shell integrals.
The equations are second order and reduce the loop order by one. We
presented several classes of infinite of integrals closed under the differential equations.
In particular, these include most integrals appearing in $\cN=4$ MHV amplitudes to two loops. 
One may speculate that the same should be true to all loop orders.
\\

We also discussed how such differential equations can be solved by making
reasonable assumptions about the boundary conditions. The explicit solution for a finite two-loop
seven-point and a divergent two-loop five-point integral was given.
The answers are relatively simple and one can hope that one
can solve for infinite series of such integrals, as is possible for the off-shell
ladder integrals \cite{Usyukina:1993ch,Broadhurst:1993ib,Isaev:2003tk}.
This would be an important step toward summing the perturbation
series in $\cN=4$ SYM. See \cite{Broadhurst:2010ds} for a study of the summation
of the off-shell latter integrals.\\

The main motivation of this study was to contribute to the determination of the all-order S-matrix of
$\cN=4$ SYM.
One may ask whether there is a short-cut by deriving differential equations for the whole amplitude,
as opposed to individual integrals. In Appendix \ref{sect-diff-integrands} we found differential operators that annihilate
(naively) the integrand of the one-loop MHV amplitude. In order to understand these differential
operators better it is important to systematically investigate distributional terms that can appear and
lead to inhomogeneous terms in differential equations.
A related open question is the link of our differential equations to the conformal symmetry of $\cN=4$ SYM.\\

We expect our differential equations to be useful for other theories as well.
Due to their nice properties it is certainly desirable to include the integrals with magic numerators
introduced in \cite{ArkaniHamed:2010kv} and discussed in \cite{Drummond:2010mb} and the present paper in the integral basis.  The
necessary numerator identities were given in \cite{Drummond:2010mb}.
Moreover, one can find differential equations of the type presented here for larger classes
of integrals. This is clear from the way the differential equations were derived: all one
needs is a certain sub-topology with ``magic'' numerators.
In the cases considered in this paper the latter appeared on top of
a pentagon topology, see Fig.~\ref{fig-key},  which made sure that the integrals are
dual conformal. However, one could drop this requirement and extend
the current method to integrals that are not dual conformal.\\

Another natural extension of our work is to use different regulators. Some of the integrals
discussed here are finite, and in the case of infrared divergences we used a mass regulator,
allowing us to stay in four dimensions, which is very natural when working with twistor variables.
However, this is not mandatory. Although the details may be slightly different, the mechanism
for finding differential equations should work with other regulators as well. E.g. in dimensional
regularisation one might expect similar equations to hold, at least to some order in $\epsilon$, where
the dimension is $4-2 \epsilon$. \\

It is interesting to obtain equations exact in $m^2$, which would allow to discuss e.g. scattering
amplitudes on the Coulomb branch of $\cN=4$ SYM. Equations exact in $m^2$ could also be useful to
study integrals appearing in realistic amplitudes involving massive particles, see e.g. \cite{Anastasiou:2006hc}.
It should be possible to derive such equations at least in some cases, perhaps by noticing that
four-dimensional massive propagators are formally identical to AdS bulk-to-boundary propagators \cite{D'Hoker:2002aw}.

\section*{Acknowledgements}
It is a pleasure to thank N. Arkani-Hamed, J. Bourjaily, F. Cachazo and A. Kotikov for useful discussions.
J.M.D. and J.M.H. are indebted to N. Arkani-Hamed for several invitations to the IAS, where
part of this work was done.
\appendix

\section{Twistor differential operators}
\label{sect-diff-integrands}

In section \ref{sect-diffeq}, we showed that all integrals appearing in
the one- and two-loop MHV amplitudes with at least
five external legs satisfy differential equations.
In order to obtain these equations it was crucial to find
differential operators that can act on a sub-integral
and reduce it to a rational function.\\

In order to find the relevant twistor differential operators we found
it useful to systematically write down all differential operators
that have of property of (naively) annihilating the integrand of
a given integral. Of course this is not equivalent to finding a 
useful differential equation for the {\it function} after integration,
but it served as a helpful starting point for our above analysis.
Here we present more details of the mechanism how differential
operators annihilate the integrand of the integrals discussed above.
The reason is two-fold: firstly, it would be interesting to understand
in more detail how the distributional term that freezes the loop integration
is generated. Secondly we expect our differential equation method
to apply to larger classes of integrals. We hope that this section will
prove a useful starting point for such investigations.\\

We do not claim that we exhaust all differential equations for a
given diagram, but we show that there is always at least one for all
diagrams appearing in the one-loop and two-loop MHV amplitudes.

\subsection{Diagrams in the one-loop amplitude}

The form of one-loop MHV amplitude presented in
\cite{ArkaniHamed:2010kv} contains sum of two-mass-hard boxes and
double pentagons with ``magic numerator''.
As was mentioned above, for convenience we use the pentagons with dashed lines, the
difference being a parity odd piece that vanishes after integration
(or is of order ${\cal O} (m^2)$ in the massive regularization).


\subsubsection{Two-mass hard diagram}

The two-mass-hard boxes as integrals in momentum twistor space are\footnote{We changed the labels from $i \to i+1$ w.r.t. to
 the main text.}
\begin{equation}
I_{2mh} =
\frac{N}{\fbr{AB12}\fbr{AB23}\fbr{AB34}\fbr{AB\,i\,i\pl1}}\,,\quad\mbox{where}\quad
N=\fbr{1234}\fbr{23\,i\,i\pl1}\,.
\end{equation}
This integrand is annihilated trivially by the first order differential
operators $O_{21}$, $O_{34}$, $O_{ii\pl1}$, $O_{i\pl1i}$ and also $O_{kj}$
for all $k$ and $j=5,\dots i-1,i+2\dots n$.

We are interested in obtaining non-trivial second order
differential equations for $I_{2mh}^i$.
Applying the operator $O_{24}$ on the
un-normalised integrand
$I_{2mh}/N$ we get
\begin{equation}
O_{24} \frac{1}{\fbr{AB12}\fbr{AB23}\fbr{AB34}\fbr{AB\,i\,i\pl1}} =
\frac{1}{\fbr{AB12}\fbr{AB34}^2\fbr{AB\,i\,i\pl1}}\,,
\end{equation}
and this is trivially annihilated by $O_{12}$ and $O_{43}$.
Similarly, we can start with the operator $O_{31}$ and then apply
$O_{12}$ and $O_{43}$. As a result, we get four differential
equations\footnote{Note that we do not include the second order
operators that include $O_{21}$ and $O_{34}$ as first order pieces.}
\begin{equation}
\tilde{O}_{1224} I_{2mh} = \tilde{O}_{4324} I_{2mh} =
\tilde{O}_{1231} I_{2mh} = \tilde{O}_{4331} I_{2mh} = 0\,,
\end{equation}
where
\begin{equation}
\tilde{O}_{ijkl} = NO_{ij}O_{kl}N^{-1}\,.
\end{equation}
In the boundary case of $i=4$ or $i=n$ (which corresponds to a one-mass
diagram) we cannot use both ``directions'' of the derivatives. So in the
first case we use only $O_{31}$ to create a massive
corner, while in the second case only $O_{24}$. There is a special
case of $n=4$ which corresponds to the zero-mass diagram. In this case we 
cannot use our construction. However, this diagram itself represents
the one-loop MHV amplitude and we will see at the end of this subsection
that there exist differential operators that annihilate naively all MHV one-loop
integrands (including the $n=4$ case).\\

For reasons that will become clear later, it will be quite useful also to
find a differential operator acting only on a limited number of twistors.
Suppose that the only operators we want to use are
$O_{k2}$, $O_{k3}$, $O_{n1}$, $O_{54}$ for any $k$. Even in that
case there exists a second-order operator, although somewhat complicated,
\begin{equation}
\left[\tilde{O}_{4354} -
\frac{\fbr{2345}}{\fbr{1234}}\tilde{O}_{1223} -
\frac{\fbr{2456}}{\fbr{2356}}\tilde{O}_{54}+\frac{\fbr{1356}\fbr{2345}}{\fbr{1234}\fbr{2356}}\tilde{O}_{23}\right]I_{2mh}
= 0 \,,
\end{equation}
and also its flipped version.
It is hard to find an explanation for
this operator (like we did for other operators above),
but its existence is important for the discussion of the
two-loop double boxes.

\subsubsection{Pentagon diagram}

The pentagon integrals in momentum twistor space are
\begin{equation}
I^{ij}_{pent} =
\frac{N\,\fbr{AB(123)\cap(i\mi1\,i\,i\pl1)}}{\fbr{AB12}\fbr{AB23}\fbr{AB\,i\mi1\,i}\fbr{AB\,i\,i\pl1}\fbr{AB\,j\,j\pl1}}\,,
\end{equation}
where $N=\fbr{2\,i\,j\,j\pl1}$.
The strategy is very similar to the two-mass hard case. We can find
many trivial first order operators that annihilate the integrand, and we can
use again the same trick of creating a massive corner. Here we have
even more freedom in the choice which propagator we want to kill, e.g.
we can first apply the operator $O_{13}$, yielding
\begin{eqnarray}
&& \hspace{-2cm} O_{13}
\frac{\fbr{AB(123)\cap(i\mi1\,i\,i\pl1)}}{\fbr{AB12}\fbr{AB23}\fbr{AB\,i\mi1\,i}\fbr{AB\,i\,i\pl1}\fbr{AB\,j\,j\pl1}} =\nonumber \\
&& =  \frac{\fbr{AB(123)\cap(i\mi1\,i\,i\pl1)}}{\fbr{AB23}^2\fbr{AB\,i\mi1\,i}\fbr{AB\,i\,i\pl1}\fbr{AB\,j\,j\pl1}}\,,
\end{eqnarray}
\normalsize
which is annihilated by the operator $O_{32}$. Finally, we can find four operators\footnote{For the boundary case when $j=i+1$ or
$j=n$ some operators are not present.} that non-trivially annihilate
$I^{ij}_{pent}$,
\begin{equation}
\tilde{O}_{3213} I^{ij}_{pent} = \tilde{O}_{1231} I^{ij}_{pent} = 0\,,
\end{equation}
and the same for $(1,2,3)\rightarrow (i\pl1,i,i\mi1)$. We see that
the operators act separately just on the left or right region.\\

Now, something interesting happens when $i=4$, i.e. when the top corner is
massless. Then both regions  $(1,2,3)\rightarrow (i\pl1,i,i\mi1)$
overlap and the equations we wrote are not valid anymore. However,
the numerator saves the situation and helps us to find another
differential equation. First, let us apply the operator $O_{23}$,
\small
\begin{align}
O_{23}
&\frac{\fbr{AB(123)\cap(345)}}{\fbr{AB12}\fbr{AB23}\fbr{AB34}\fbr{AB45}\fbr{AB\,j\,j\pl1}} = \nonumber
\\&=
\frac{\fbr{AB(123)\cap(245)}}{\fbr{AB12}\fbr{AB23}\fbr{AB34}\fbr{AB45}\fbr{AB\,j\,j\pl1}}
-
\frac{\fbr{AB24}\fbr{AB(123)\cap(345)}}{\fbr{AB12}\fbr{AB23}\fbr{AB34}^2\fbr{AB45}\fbr{AB\,j\,j\pl1}}\nonumber\\
&= \frac{\fbr{1234}}{\fbr{AB12}\fbr{AB34}\fbr{AB\,j\,j\pl1}}\,,
\end{align}
\normalsize
where we used $\fbr{AB(123)\cap(234)} = \fbr{AB23}\fbr{1234}$. The
term we obtained is annihilated by $O_{12}$ and $O_{34}$. Similarly,
we can start with the operator $O_{43}$ and repeat the above calculation.
In total we get four differential equations,
\begin{equation}
\tilde{O}_{1223}I_{pent}^{4j} = \tilde{O}_{3423}I_{pent}^{4j} =
\tilde{O}_{5443}I_{pent}^{4j} = \tilde{O}_{3243}I_{pent}^{4j} = 0\,.
\end{equation}
Here there are no boundary cases for $j=5$ or $j=n$.
\\

In addition to these operators, we can also find two more special
operators that annihilate the integrand of the pentagon. Like in
the two-mass hard case they also contain only a restricted number of
derivatives, but their forms are much nicer,
\begin{equation}
\left[\sum_{k=2}^{i-2}\tilde{O}_{ki\mi11k} + O_{1i\mi1}\right]
I^{i,j}_{pent} = \left[\sum_{k=4}^{i}\tilde{O}_{k3i\pl1k} +
O_{3i\pl1}\right] I^{i,j}_{pent} = 0\,.
\end{equation}
We can see that two of equations for $i=4$ were exactly of this
type.

\medskip

We completed the picture of differential operators for all diagrams
in one-loop MHV amplitude. Of course, the one-loop MHV amplitude is known
analytically, but this result will be extremely useful in analyzing
the diagrams contributing to the two-loop MHV amplitude.

\subsubsection{Differential equations for the one-loop MHV amplitude}

So far, we analyzed the one-loop amplitude giving the equations for
all diagrams contributing to it. However, there exists an equation
for the integrand of the whole amplitude.
In fact, for the $n$-point amplitude there exist $n^2$ differential equations of the
form
\begin{equation}
Q_{ij} I_{MHV}^{1-loop} = 0\qquad\qquad i,j=1,\dots n
\end{equation}
where
\begin{equation}
Q_{ij} = \sum_{k=1}^n O_{ik}O_{kj} - (n-2)\,O_{ij}\,.
\end{equation}
Note that the structure of operators is quite similar to
that which we found for the pentagons at the end of the last subsection.

\subsection{Diagrams in the two-loop amplitude}

We start with the form of two-loop MHV amplitude given in
\cite{ArkaniHamed:2010kv} and reviewed in section \ref{sect-moti}.
It contains three topologies:
double-boxes, penta-boxes and double-pentagons.


We will analyze each topology separately. Naively, they look like
one-loop diagrams (two-mass hard integrals and pentagons) glued together,
and we will see that in some sense they really are.\\

We want to find a  second-order differential equation for the integrands of these
diagrams, keeping in mind our  motivation that acting with the correct second-order differential
operator on a two-loop diagram can decrease the loop level by one).
Therefore, it would be very natural to act just on a one-loop
sub-diagram, which is exactly what we are going to do. We will see
that there is close link with the operators found in the last
subsection.

\subsubsection{Double-boxes}

The integrand for the general double-box appearing in the amplitude
\begin{equation}
\hspace{-0.1cm}I_{double-box}^{i} =
\frac{\fbr{1234}\fbr{i\mi1\,i\,i\pl1\,i\pl2}\fbr{23\,i\,i\pl1}}{\fbr{AB12}\fbr{AB23}\fbr{AB34}\fbr{ABCD}
\fbr{CD\,i\mi1\,i}\fbr{CD\,i\,i\pl1}\fbr{CD\,i\pl1\,i\pl2}}\,,
\end{equation}
can be rewritten in a very suggestive form
\begin{equation}
\frac{\fbr{1234}\fbr{23\,i\,i\pl1}}{\fbr{AB12}\fbr{AB23}\fbr{AB34}\fbr{AB\,i\,i\pl1}}\cdot
\frac{\fbr{i\mi1\,i\,i\pl1\,i\pl2}\fbr{AB\,i\,i\pl1}}{\fbr{ABCD}\fbr{CD\,i\mi1\,i}\fbr{CD\,i\,i\pl1}\fbr{CD\,i\pl1\,i\pl2}}\,.
\end{equation}
We see that the double box can be written in a factorised form as a
product of two-mass hard diagrams (note that they are properly
normalised), where the right CD sub-diagram contains twistors A,B as
external, but the left AB sub-diagram depends only on external twistors.
If $5<i<n-1$ (both internal corners are massive), then there is no
overlap between AB and CD sub-diagrams and the AB part is
annihilated by the same operators as the normal 2-mass-hard integrand,
\begin{equation}
\tilde{O}_{1224} I_{2box} = \tilde{O}_{4324} I_{2box} =
\tilde{O}_{1231} I_{2box} = \tilde{O}_{4331} I_{2box} = 0
\end{equation}
where here by the normalisation $N$ we mean the normalisation of
left sub-diagram $N=\fbr{1234}\fbr{23\,i\,i\pl1}$. If $i=4,5,n-1,n$
just
half of the operators survive (we still have one massive corner).\\

For $n=4,5,6$ we have diagrams that do not contain any
massive corners. 
However, the
diagrams for $n=5$ and for $n=6,i=5$ satisfy differential equations
and here we use the complicated forms that we found in the end of
the discussion of the two-mass hard boxes.
Thanks to restrictions we made on
the first-order operators for the construction of the second-order
operator, these operators act just on one half of the double-box
integrand even if the indices overlap (which exactly happens for
$i=5$ or $i=n-1$). The only constraint is that at least one
external line (which can be massless and just on one side of the diagram)
must be attached at the point were the two one-loop diagrams are glued together.
Then the equations are exactly the same as we found for the two-mass hard case,
since we act on just half of the diagram.


\subsubsection{Pentagon-boxes}

As in the previous case, we can write the integrand for the
pentagon-box appearing in the two-loop MHV amplitude,
\footnotesize 
\begin{equation}
\hspace{-0.145cm}I^{penta-box}_{n;i,j} =
\frac{\fbr{AB(123)\cap(i\mi1\,i\,i\pl1)}\fbr{2\,i\,j\,j\pl1}
\fbr{j\mi1\,j\,j\pl1\,j\pl2}}{\fbr{AB12}\fbr{AB23}\fbr{AB\,i\mi1\,i}\fbr{AB\,i\,i\pl1}\fbr{ABCD}\fbr{CD\,j\mi1\,j}\fbr{CD\,j\,j\pl1}\fbr{CD\,j\pl1\,j\pl2}}\,,
\end{equation}
\normalsize
in a factorised form,
\footnotesize 
\begin{align}
&
\frac{\fbr{AB(123)\cap(i\mi1\,i\,i\pl1)}\fbr{2\,i\,j\,j\pl1}}{\fbr{AB12}\fbr{AB23}\fbr{AB\,i\mi1\,i}\fbr{AB\,i\,i\pl1}\fbr{AB\,j\,j\pl1}}\cdot
\frac{\fbr{j\mi1\,j\,j\pl1\,j\pl2}\fbr{AB\,j\,j\pl1}}{\fbr{ABCD}\fbr{CD\,j\mi1\,j}\fbr{CD\,j\,j\pl1}\fbr{CD\,j\pl1\,j\pl2}} = \nonumber\\
&
\frac{\fbr{AB(123)\cap(i\mi1\,i\,i\pl1)}\fbr{CD2\,j}}{\fbr{AB12}\fbr{AB23}\fbr{AB\,i\mi1\,i}\fbr{AB\,i\,i\pl1}\fbr{ABCD}}\cdot
\frac{\fbr{2\,i\,j\,j\pl1}\fbr{j\mi1\,j\,j\pl1\,j\pl2}}{\fbr{CD2\,j}\fbr{CD\,j\mi1\,j}\fbr{CD\,j\,j\pl1}\fbr{CD\,j\pl1\,j\pl2}}\,.
\end{align}
\normalsize
We see that there are two choices, either the pentagon or the two-mass-hard is
the ``proper'' one-loop subdiagram.


If we use the first choice, then the the two-loop diagram is
annihilated by all operators that annihilate its pentagon one-loop
sub-diagram, and the same is true for the second choice, where we use the
differential operators for the two-mass-hard diagram.
If the indices do not overlap, we can just act with the
operators that we have for pentagons, resp. two-mass hard and we
annihilate the integrand. In the special case of $j=i+1$ or $j=n-1$
we would repeat the same discussion as in the case of double-boxes.
Here we use the more complicated forms that we derived in the end of
the two-mass hard and pentagon sections that can be used even if the
indices overlap. Therefore, all pentagon-box integrands satisfy second-order
differential equations.

\subsubsection{Double-pentagons}

Finally, we can write the integrand for the double-pentagons,
\footnotesize 
\begin{equation}
\frac{\fbr{AB(123)\cap(\ell\mi1\,\ell\,\ell\pl1)}\fbr{CD(i\mi1\,i\,i\pl1)\cap(j\mi1\,j\,j\pl1)}\fbr{2\,\ell\,
i\,j}}{\fbr{AB12}\fbr{AB23}\fbr{AB\,\ell\mi1\,\ell}\fbr{AB\,\ell\,\ell\pl1}\fbr{ABCD}\fbr{CD\,i\mi1\,i}
\fbr{CD\,i\,i\pl1}\fbr{CD\,j\mi1\,j}\fbr{CD\,j\,j\pl1}}\,,
\end{equation}
\normalsize
in a factorised form,
\footnotesize 
\begin{align}\nonumber
\frac{\fbr{AB(123)\cap(\ell\mi1\,\ell\,\ell\pl1)}\fbr{2\,\ell\,
i\,j}}{\fbr{AB12}\fbr{AB23}\fbr{AB\,\ell\mi1\,\ell}\fbr{AB\,\ell\,\ell\pl1}\fbr{AB\,i\,j}}
  \frac{\fbr{CD(i\mi1\,i\,i\pl1)\cap(j\mi1\,j\,j\pl1)}\fbr{AB\,i\,j}}{\fbr{ABCD}\fbr{CD\,i\mi1\,i}\fbr{CD\,i\,i\pl1}\fbr{CD\,j\mi1\,j}\fbr{CD\,j\,j\pl1}}\,.
\end{align}
\normalsize
We again see two correctly normalised pentagons, and
one of them is a
proper pentagon on which we can apply differential operators. The
discussion would be identical to the previous cases, one just uses the
operators found in the pentagon section which are guaranteed to
annihilate also the double-pentagon integrand.

\section{Five-point pentagon and penta-box integrals}
\label{app-5pt}

As an explicit example of the differential equations for regulated integrals of section \ref{sect-diffeq},
we give the result for the five-point pentagon and penta-box integrals at  one and two loops.
Let us start with the one-loop pentagon integral. Without loss of generality, we choose it
in the orientation
\begin{equation}
F^{\rm pent}_{5;5,1,3} = \int \frac{d^{4}Z_{AB}}{i \pi^2} \, \frac{ \fbr{2345} \fbr{4512} \fbr{AB13} }{\fbr{AB51} \fbr{AB12} \fbr{AB23} \fbr{AB34} \fbr{AB45} } \,,
\end{equation}
and we recall the reader that because of infrared divergences we also include $+m^2$ terms in the definition of the propagators, see section \ref{sect-moti}.
Introducing Feynman parameters, we obtain
\begin{equation}
F^{\rm pent}_{5;5,1,3} = -2\, x_{35}^2 x_{25}^2 x_{14}^2 \int_{0}^{\infty} \frac{ d\alpha_{i} \alpha_{5} \delta(\sum_{i=1}^{5} \alpha_{i} -1) }{(\sum_{i=1}^{5} \alpha_{i} \alpha_{i+2} x_{i,i+2}^2 + m^2 )^3}  + \cO(m^2)\,.
\end{equation}
The small $m^2$ expansion of this integral can be conveniently obtained using Mellin-Barnes methods.
The answer depends on three variables,
$y_1=x^2_{35}/x^2_{13}, y_2 = x^2_{25}/x_{24}^2 , y_{3}= x^2_{14}/m^2$,
and can be written as
\begin{equation}\label{magic-5ptpentagon-nice}
F^{\rm pent}_{5;5,1,3}  =  -\frac{1}{2}  \log^2(y_1 y_2 y_3) -2 \, {\rm Li}_{2}( 1 - y_1) -2 \, {\rm Li}_{2}( 1 - y_2) + \frac{\pi^2}{6}  +\cO(m^2) \,.
\end{equation}
We can also compute the corresponding two-loop penta-box integral.
In the nomenclature of section \ref{sect-moti}, it is
\begin{eqnarray}
F^{\rm penta-box}_{5;5,1,3} \hspace{-0.3cm} &=& \hspace{-0.3cm}  \int  \frac{d^{4}Z_{AB} d^{4}Z_{CD} \, (i \pi^2)^{-2}\, N \fbr{CD13}  }{ \fbr{AB34} \fbr{AB45} \fbr{AB51} \fbr{ABCD} \fbr{CD51}\fbr{CD12}\fbr{CD23}\fbr{CD34}} \,,\phantom{space}
\end{eqnarray}
where $N=\fbr{3451} \fbr{2345} \fbr{4512}$.
It evaluates to
\begin{eqnarray}
F^{\rm penta-box}_{5;5,1,3}  = h(y_1 y_2 y_3 ) - 2 H_{0}(y_1 y_2 y_3 ) \left[ f(y_1 ) + f( y_2 ) \right] +  \left[ g(y_1 ) + g( y_2 ) \right] +\cO(m^2) \,,
\end{eqnarray}
where
\begin{eqnarray}
h(x) &=& -\frac{7}{40} \pi^4 + 2 \zeta_{3} H_{0}(x) - \frac{1}{2} \pi^2 H_{0,0}(x)-H_{0,0,0,0}(x) \,,\\
g(x) &=&
-\frac{2}{3} H_{2}(x) + 2 H_{0}(x) H_{2,0}(x) -
 4 H_{3,0}(x) - 2 H_{2,0,0}(x) - 4 H_{2,1,0}(x)\,, \\
 f(x) &=& H_{0,1,0}(x)\,.
\end{eqnarray}
The integrals satisfy
\begin{equation}
y_{1} \frac{\partial}{\partial y_{1}}  y_{2} \frac{\partial}{\partial y_{2}}  F^{\rm pent}_{5;5,1,3} =  - 1 + \cO(m^2) \,,
\end{equation}
and
\begin{equation}
y_{1} \frac{\partial}{\partial y_{1}}  y_{2} \frac{\partial}{\partial y_{2}}  F^{\rm penta-box}_{5;5,1,3}
 =   F^{\rm pent}_{5;5,1,3} + \cO(m^2) \,.
\end{equation}
One can check that equations (\ref{divpent1loop}) and (\ref{divpent2loop}) are indeed satisfied.

\bibliographystyle{nb.bst}
\bibliography{loops}

\end{document}